\documentclass[aps,showpacs,preprintnumbers, superscriptaddress, amsmath,amssymb,nofootinbibt,referee, twocolumn]{revtex4-2}
\usepackage{amsmath}
\usepackage{eurosym}
\usepackage{amssymb}
\usepackage{graphicx}
\usepackage{color}

\def\be{\begin{equation}}
\def\ee{\end{equation}}
\def\bea{\begin{eqnarray}}
\def\eea{\end{eqnarray}}

\begin{document}

\title{Warm inflation with non-comoving scalar field and radiation fluid}
\author{Tiberiu Harko}
\email{tiberiu.harko@aira.astro.ro}
\affiliation{Astronomical Observatory, 19 Ciresilor Street, 400487 Cluj-Napoca, Romania,}
\affiliation{Department of Physics, Babes-Bolyai University, Kogalniceanu Street,
Cluj-Napoca 400084, Romania,}
\affiliation{School of Physics, Sun Yat-Sen University, Xingang Road, Guangzhou 510275, People's
Republic of China}
\author{Haidar Sheikhahmadi}
\email{h.sh.ahmadi@gmail.com; h.sheikhahmadi@ipm.ir}
\affiliation{Center for Space Research, North-West
University, Mafikeng, South Africa}
\affiliation{School of Astronomy, Institute for Research in
Fundamental Sciences (IPM),  P. O. Box 19395-5531, Tehran, Iran}

\begin{abstract}
We consider a warm inflationary scenario in which the two dominant matter
components present in the early Universe, the scalar field, and the radiation fluid,
evolve with different four-velocities. This cosmological system is
mathematically equivalent to a single anisotropic fluid, evolving with a four-velocity
that is a function of the two independent fluid four-velocities. Due to the presence of the anisotropic physical parameters,  the overall cosmological evolution is also anisotropic. We derive the gravitational field equations of the noncomoving scalar field-radiation mixture for a Bianchi type I geometry. By considering that the decay of the scalar field is accompanied by a corresponding radiation generation, we formulate the basic equations of the warm inflationary model in the presence of two noncomoving components. By adopting the slow roll approximation, we perform a detailed comparison of the theoretical predictions of the warm inflationary scenario with noncomoving scalar field and radiation fluid with the observational data obtained by the Planck satellite, by investigating both the weak dissipation and strong dissipation limits. Constraints on the free parameters of the model are obtained in both cases. The functional forms of the scalar field potentials compatible with the noncomoving nature of warm inflation are also derived.
\end{abstract}
\date{\today}
\maketitle
\tableofcontents

\section{Introduction}

The Planck data of the 2.7 full sky survey, recently released, \cite{1,2,3,4,5,pl2018a,pl2018b}
have indicated a number of interesting aspects, whose account will certainly
require a deep modification in our present day understanding of the Universe. The current
observations have determined the physical properties of the Cosmic Microwave Background Radiation (CMB) with a
remarkable precision. One of the most substantial results of the Planck satellite
mission is the important result that the best-fit Hubble constant has the numerical value $H_0 = 67.4\pm
1.2 $ km s$^{-1}$Mpc$^{-1}$, while the dark energy density parameter is equal to $\Omega
_{\Lambda } = 0.686 \pm 0.020$, with the other fundamental cosmological parameter,  the matter density parameter given by $\Omega _M =
0.307 \pm 0.019$.

Generally, the Planck data have confirmed the
basic theoretical aspects of the standard $\Lambda $CDM ($\Lambda $Cold Dark Matter) paradigm. According to the $\Lambda $CDM model, the main material composition of the Universe
can be reduced to two basic components only, called dark energy, and dark matter, respectively \cite%
{PeRa03, Padh}. The late-time cosmic acceleration of the Universe \cite%
{acc}, whose observation has led to a drastic change in our views of the Universe,  can be outstandingly be explained by assuming either the existence of a fundamental geometrical(or perhaps physical) parameter, the cosmological constant $\Lambda $ \cite{1z}. From a geometric point of view  $\Lambda$ would represent an intrinsic
curvature of the space-time. Alternatively, one could interpret observations by postulating the existence of the dark energy, an assumed fluid component corresponding to a zero-point-energy of a physical field that permeates the whole Universe. Dark energy should
 resemble  a cosmological constant during the late phases of the
cosmological expansion \cite{PeRa03, Padh, revn1, Sheikhahmadi:2014rka, revn2,revn3,revn4}. Presently,  one of the important dark energy scenarios is
the so-called \textit{quintessence} model
\cite{quint}-\cite{quint4}, in which dark energy is generated by a scalar particle $\phi $. For other proposals for dark energy, in which the dynamical equation of state is realized by a scalar field, one can refer to k-essence \cite{{6wN},{6awN},{6bwN}}, tachyon {\cite{7wN, Rasheed:2020syk}}, phantom \cite{{8wN},{8awN},{8bwN}}, quintom\cite{{9wN},{9awN},{9bwN}}, chameleon  
\cite{10wN}-\cite{19aawN},
and Chaplygin gas  \cite{Chap,Chap1} models have also been investigated. Assuming extra-dimensional cosmological effects, or modifying gravity at astrophysical, galactic or extra-galactic scales,  could also lead to an account of the recent acceleration
of the Universe 
\cite{mod1}-\cite{mod8}. Moreover,its has been suggested that scalar fields, or other long range coherent physical fields
minimally or nonminimally coupled to gravity, can also be considered as promising dark matter candidates
\cite{OvWe04, revdm1,revdm2, revdm3}.

The basic paradigm of the present day cosmology about the very early Universe is represented
by the inflationary theory, initiated in \cite{infl1}. The basic idea of inflation is the existence in the very early Universe of a scalar field $\phi$, characterized by a self-interaction potential $V(\phi)$, and having a physical energy density $\rho _{\phi}$, and  a pressure $p_{|phi}$, respectively \cite{infl2}. The early inflationary models were based on the premise that at $\phi = 0$ the scalar field potential attains a local minimum. This behavior can be explained by the  supercooling after a phase transition. Subsequently, an exponentially expanding, de Sitter type phase describes the evolution of the Universe. However,  in this initial cosmological approach, called the old inflationary scenario, there is no graceful exit from the eternally expanding,  de Sitter type inflationary phase.
Many inflationary models, with the explicit objective of solving the graceful exit problem, have been proposed, including the so-called new and chaotic
inflationary approaches  \cite{infl3,infl4,infl5}. It is important to note that each
of these models have their specific, and generally still unsolved,  theoretical problems. For recent
presentations of different aspects of inflationary cosmology see \cite{infl7,infl8}.

During inflation the de Sitter type expansion of the scale factor results in a homogeneous, isotropic but matterless Universe, in which all initial matter components have diminished to near zero.
Hence, in order to explain the present day composition of the Universe, one must assume that radiation and the basic elementary particles have been created at the end of inflation, in a cosmological epoch known as  reheating. During reheating matter (existing initially mostly in the form of electromagnetic radiation) was created through the transfer
of energy from the scalar field triggering the inflationary evolution to the elementary particles. Initially, the reheating theoretical model was  developed by using the theoretical formalism of the
new inflationary scenario \cite{infl3}, and later on extended in many studies \cite{reh1, reh2, reh3}.  The essential idea of reheating can be represented in the following way. Once the de Sitter type accelerated expansion of the Universe did end, the scalar field prompting inflation did reach its minimum value. Then it began to oscillate around the minimum of the potential, and subsequently it disintegrated into
matter, in the form of a radiation (photon) gas, and of some Standard Model elementary particles. Due to the
interaction of these components,  the early Universe  finally reached a state of thermal equilibrium,
 characterized by a single temperature $T$.

One of the approaches extensively used to investigate reheating is the phenomenological model developed in
\cite{infl4}. The basic idea of the formalism in \cite{infl4} is the addition of a particular decay term in the Klein-Gordon equation that describes the time evolution of the scalar field.  This term is also
incorporated as a growth term in the balance equation for the energy density of the particles newly created by the scalar field. If reasonably chosen, the loss/gain terms can give a good description of the reheating process that did follow the adiabatic supercooling of the Universe during the de Sitter type expansion of the inflationary era. Therefore,
in this simple reheating model, an interacting two component fluid model, consisting of a mixture between the scalar field, and ordinary matter (radiation), can explain the present day chemical composition of the Universe. Thus, the entire complex transition process between
the scalar field and the radiation component can be described through this simple phenomenological approach, once the functional form of the friction term, describing the time decay of the scalar inflaton field, is given. The same friction term also acts as a source term for the
newly created matter particle, usually assumed to be photons. For investigations of the diverse cosmological and physical aspects of the post inflationary
reheating dynamics see \cite{reh4}-\cite{reh15}.
Detailed reviews on the reheating phase that followed inflation can be found in \cite{rehr1} and \cite{rehr2}, respectively.

Despite its remarkable theoretical success the standard reheating model is plagued by a number of
problems. One important question is related to the perturbative description of the decay
width of the scalar field, which can describe the decay only very close to the minimum of the scalar field potential. Such a description is not valid  during
slow-roll inflation. Another problem is that the effects of the high finite temperature of the scalar field, and of the cosmic environment,  can also significantly increase the rate at which the inflaton field dissipates
its energy to the newly created particles \cite{rehr1,rehr2}.

Moreover, it is natural to assume that the scalar field driving inflationary expansion could have
been nonminimally  coupled to the other components of the cosmological fluid present in the very
early Universe. Consequently, the scalar field could have dissipated its energy
during the inflationary phase itself, thus warming up the Universe without the need of a reheating phase. This view of the inflation phase is called warm inflation, and it was proposed for the first time in \cite{w1,w2}. According to the warm inflationary scenario, during the accelerated inflationary expansion phase dissipative effects and particle creation processes can create a thermal bath, consisting of a radiation fluid.
In one of the first warm inflationary models \cite{w3} it was proposed that in the inflationary scenarios
the physical and cosmological parameters could be randomly distributed. This hypothesis led to the development of the so-called distributed-mass models \cite{w3a, w4,w5,w6,w7}, also investigated in relation to string theory. Warm inflation represents presently a very active
field of research, and it certainly represents an attractive and interesting  alternative to the standard cold
inflation/reheating scenarios. Different aspects of the cosmological evolution in the warm inflationary models have been
investigated in great detail in \cite{w8}-\cite{w57}.


The Planck observational data also point towards the possible existence of some tension between the fundamental principle of
the $\Lambda$CDM paradigm and the cosmological observations. For example, after combining the WMAP polarization data with the Planck data,  the index of the power spectrum is  found to be
$n_s=0.9603\pm0.0073 $ \cite{2,3}, at the pivot scale $k_0 = 0.05$ Mpc$^{-1}$. This value excludes the exact scale-invariance ($n_s = 1$) at more than $%
5\sigma $. Moreover, the joint constraints on the tensor to scalar ratio $r$ and $n_s$ are able to significantly restrain inflation models. On the other hand, one should note that inflationary models with a power-law potential of the form $\phi ^4$ cannot
provide a legitimate number of $e$-folds (between 50-60) in the restricted
space of $r-n_s$ at around a $2\sigma $ level. Hence the precise observations of the Cosmic Microwave Background Radiation
permit the testing of some basic predictions of inflationary models  on primordial
fluctuations, such as Gaussianity and scale independence \cite{1}-\cite{pl2018b}.


An even more interesting result that emerged from the explorations of the background geometry and topology of the Universe indicated that a Bianchi type geometric structure, which may correspond to a homogeneous but anisotropic geometry of the Universe, could explain some of the large-scale anomalies detected by the Planck satellite \cite{5}.  By using the Planck data, a Bayesian search
for an anisotropic and homogeneous Bianchi type $\mathrm{VII_h}$ geometry was carried out in \cite{5}. It turns out that in a non-physical setting, with the Bianchi
parameters decoupled from the standard cosmology, the observational data favor a Bianchi component with a Bayes factor of at least 1.5 units of
log-evidence. On the other hand, in the physically motivated cosmological configuration, where the Bianchi parameters of the anisotropic model are fitted simultaneously with the standard $\Lambda$CDM cosmological
parameters, no indication for a Bianchi $\mathrm{VII_h}$ geometry was detected \cite{5}.

The statistical isotropy, characterizing the large scale structure of the Universe, is an important prediction of standard cosmology, based on the hypothesis that inflation eliminates classical (or even quantum) anisotropy. This result is also strongly supported by the
cosmic no-hair conjecture. However, a number of recent observations of the large scale structure of the Universe have questioned the basic principles of homogeneity and
isotropy, on which modern cosmology is built up  \cite{an, ann}. Some of the recent observations, not related to the study of the CMB, and pointing towards possible anisotropies in the Universe are obtained from the investigations of Type Ia Supernovae, of the X-ray background, the distribution of the optical and infrared galaxies, and the observation of some peculiar velocities of the galaxy clusters \cite{ann} may raise some concerns about the absolute validity of the principle of isotropy. In \cite{ann} the directional behavior of the X-ray luminosity-temperature relation $L_X-T$ of galaxy clusters was investigated. The measurement of the luminosity depends on the considered cosmological model via the luminosity distance $D_L$. On the other hand, the temperature can be determined astrophysically without any cosmological assumptions. It was found that the behavior of the $L_X-T$ relation strongly depends on the direction in the sky, an important result consistent with
previous investigations. Strong anisotropies were detected at a $\geq 4\sigma $ level. From the study of a sample of 142,661 quasars, with the data extending beyond the post-inflationary causality scales, compelling spatially correlated systematic effects that can mimic anisotropy on cosmological scales were found in \cite{annw}. When interpreted together with the recent Planck results, these powerful
observational evidences indicate that the possibility of the presence of a peculiar large scale anisotropy in the Universe, of cosmological origin, cannot be ruled out \textit{a priori}.

The intriguing circumstance that the geometry of the Universe may not be fully described by the standard
Friedmann-Lemaitre-Robertson-Walker (FLRW) metric was analyzed, from
different geometrical and physical perspectives, in \cite{TK}-\cite{an9}.

In particular, so-called extended FLRW models, describing cosmological evolutions in a background anisotropic Bianchi geometry that still evolves isotropically, and can be uniquely related with a standard FLRW model having the same evolutionary history, were investigated in \cite{an9}.  It was found that geometrical and matter anisotropies likely cancel out each other through a dynamical process, and that, under rather general conditions, the expansion is asymptotically isotropic.

Hence, it turns out that there is a significant theoretical and empirical indication for the possible  existence of the large scale cosmological anisotropies. However, the physical origin of anisotropy is still unknown, with the most favored explanations related to the deviations of the primordial fluctuations from
isotropy \cite{1}. Moreover, up to now  no convincing and theoretically well motivated physical process that could lead to such interesting deviations from isotropy has been proposed.

A possible physical mechanism for the generation of the anisotropies in the Universe was suggested in \cite{Ha_Lo_2013}, and it is based on the idea that the two major matter components of the Universe, dark energy and dark matter, respectively, evolve with different four-velocities. Such a cosmological configuration with two fluids is identical to a cosmological system consisting of a single anisotropic fluid \cite{Le, Le1,Ol,He}. This anisotropic fluid expands with a four-velocity obtained through the (nonlinear) combination of the  four-velocities of the two fluids. Therefore, if a slight difference between the four-velocities of the dark matter and dark energy is generated due to some specific coupling processes, the Universe would achieve some anisotropic properties. Consequently, on cosmological scales its geometry will depart from the standard FLRW one. This conclusion is supported by a number of present day observations. Along the same line of thought in \cite{Cem}  it was pointed out that there is no a priori reason to require for the dark component of the Universe a reference frame comoving with ordinary matter. The consequences of relaxing this assumption were investigated through the study of the cosmological behavior of non-comoving fluids. From the point of view of the cosmologically observable effects, important  modifications in the density-lensing potential and density-velocity cross-correlation spectra were found. The deviations from the noncomoving motion of the components of the Universe generate modifications from statistical isotropy having a dipolar structure.

A dark matter model, in which dark matter is described by a two-component thermodynamic system, with no interaction between the particles of different species,  was considered in \cite{Ha}. In this approach each distinct dark matter component has its own four-velocity, and therefore no unique comoving frame can be adopted.  The properties of such a system were further investigated in \cite{Ha1}, under the supplementary hypothesis that the two independent dark matter components consist of pressureless, noncomoving fluids, having different four-velocities. Again, the standard mathematical and physical result that the two-fluid noncomoving dark matter distribution can be described as a single anisotropic fluid, with vanishing tangential pressure, and non-zero
radial pressure, was used to obtain the dynamical behavior of the system. The interesting likelihood that different rest frames for dark matter
and dark energy could exist has been also studied in \cite{Mar,Mar1,Mar2,Mar3}. The presence in the Universe of large scale bulk flows could represent some important indication for the presence
of moving dark energy in the very early cosmological era when the decoupling of photons from matter took place.

It is the purpose of the present paper to extend the previous studies of the possibility of the noncomoving motion of the cosmological components to the very early Universe, and, more exactly, to the inflationary era. More specifically, we will consider the warm inflationary scenario, in which the early Universe is modeled as an interacting two component fluid. In the standard warm inflationary scenario a basic (but not exactly justifiable) assumption is that all matter constituents of the early Universe move in the same rest frame, and  with the same four-velocity. This idea represents a basic  hypothesis of the warm inflation theory. However, it allows to adopt a frame that is comoving
with both the scalar field and the matter (radiation) constituents, thus allowing to chose all the components of the four velocities
$u^{\mu }$ as $u^{\mu }=(1,0,0,0)$. An important consequence of the assumption of the comoving motions is that
the global thermodynamic parameters of the inflationary Universe can be obtained by simply
summing up the individual thermodynamic parameters of the component fluids. Hence, in the framework of
the standard warm inflationary model the total energy density of the early Universe $\rho $ can be obtained as $%
\rho =\sum _{i=1}^{n}{\rho _i}$, where $\rho _i$, $i=1,2,..,n$ are the
energy densities of the individual constituents (generally radiation and scalar field, respectively).

However, there is no fundamental observational or  physical principle that would
require that all matter and energy constituents in the early Universe must have the same
four-velocity.  Hence there is no a priori reason  to describe the dynamics of the inflationary Universe in a
single frame, comoving with the scalar field as well as the matter constituents. In the following we
will investigate the warm inflationary model by assuming that the very early Universe can be described as a
mixture of two interacting perfect fluids, namely scalar field and radiation, respectively, possessing different four-velocities. Therefore the two components noncomoving early Universe becomes identical, from a mathematical as well as physical point of view, to a single anisotropic fluid, as already pointed out in  \cite{Ha_Lo_2013, Le, Le1, Ol, He, Ha,Ha1}. Therefore, if a slight difference between the four-velocities of the scalar field and radiation does exist, the very early Universe would achieve some anisotropic properties, and its geometry, as well as its expansionary evolution, will not be anymore the standard FLRW one.

 In our present study we adopt the basic assumption that the scalar field and the radiation fluid have distinct four velocities. By using a rotation in the velocity space, we can transform the energy-momentum tensor of the noncomoving two fluid configuration to the standard form of anisotropic fluids. Due to this procedure the thermodynamic parameters (energy densities and pressures of the scalar field and of the radiation) of the warmly inflating  Universe are described in terms of a
single cosmological fluid, represented from a physical point of view  by an anisotropic energy-momentum tensor, from which all the physical properties of the system can be derived. The energy density of the single anisotropic cosmological fluid dominating the early Universe is larger than the sum of the energy densities
of the scalar field and of the radiation, respectively, and it contains a supplementary term due to the anisotropy induced by the noncomoving motion. For the very early Universe  we assume the simplest case of a Bianchi type I geometry, which is a consequence of the noncomoving expansion of the inflaton scalar field, and of the radiation fluid, respectively.  For this two fluid anisotropic system we derive the anisotropic gravitational field equations in their general form, as well as the generalized Klein-Gordon equation describing the cosmological evolution of the inflaton scalar field.

Once the general formalism is developed, we implement the idea of the warm inflation by introducing the decay terms for the scalar field and the radiation in their respective evolution equations. The splitting of the total energy conservation equation of the two-fluid cosmological system gives the decay equations for the scalar field, and for the radiation creation, respectively. As opposed to the standard warm inflationary scenario, the source terms for scalar field decay and radiation creation are not equal, with the radiation balance equation containing an anisotropic term proportional to the Hubble factor along the $z$ direction. We investigate in detail the slow roll approximation of the model, as well as its consistency with observations, in the two standard limits usually considered in the literature, the weak dissipation and the strong dissipation limits, respectively. Then the theoretical predictions of the warm inflationary scenario with noncomoving scalar field and radiation fluid are compared in detail with the observational data obtained by the Planck satellite, and  a number of  constraints on the free parameters of the model are obtained. The functional forms of the scalar field potentials, compatible with the noncomoving nature of warm inflation are also obtained by considering some distinct choices for the scale factors.

The present paper is organized as follows. The reformulation of the energy-momentum tensor of the scalar field-radiation fluid two component
system as energy-momentum tensor of a single effective anisotropic fluid is presented in
Section~\ref{sect2}. The warm inflationary model with noncomoving scalar field and radiation is introduced in Section~\ref{sect3}, where the balance equations for the scalar field and radiation are obtained for a Bianchi type I geometry. The slow roll approximation is investigated in Section~\ref{sect4}. We discuss and conclude our results in Section~\ref{sect5}.

\section{Non-comoving
scalar-field-radiation fluid cosmological models}\label{sect2}

We assume in the following that during its inflationary phase the early Universe consisted of a mixture of two fluid
constituents: the first component is represented by a scalar field $\phi$, with energy density and pressure $\rho
_{\phi}$ and $p_{\phi}$, and four-velocity $u_{\phi}^{\mu }$, respectively. The second component is
a radiation fluid, characterized by the thermodynamical parameters $\rho _{rad}$
and $p_{rad}$, respectively, and four-velocity $u_{rad}^{\mu }$. The
dynamical evolution of the system can then be obtained from the variational
principle
\begin{equation}
S=\int{\left[\frac{M_{Pl}^2}{16\pi}R+\frac{1}{2}\nabla _{\mu}\phi \nabla
^{\mu}\phi -U(\phi)+L_{rad}\right]\sqrt{-g}d^4x},
\end{equation}
where $M_{Pl}=G^{-1/2}$ is the Planck mass, $U(\phi)$ is the
self-interaction potential of the scalar field, and $L_{rad}$ is the
radiation Lagrangian. The Lagrangian density of the scalar field is $%
L_{\phi}=(1/2)\nabla _{\mu}\phi \nabla ^{\mu}\phi -U(\phi)$. By $\nabla
_{\mu}$ we denote the covariant derivative with respect to the metric.

\subsection{Four-velocities and energy - momentum tensors}

Now by varying the total action with respect to the metric and the scalar
field $\phi$, we obtain the energy-momentum tensor of the system as
\be
T^{\mu
\nu}=T^{\mu \nu}_{(\phi)}+T^{\mu \nu}_{(rad)},
\ee
 where
 \be
 T^{\mu
\nu}_{(\phi)}=\nabla ^{\mu}\phi \nabla ^{\nu}\phi-L_{\phi}g^{\mu \nu}.
\ee

We would like now to reformulate the energy-momentum of the scalar field in
a form similar to that of the energy-momentum tensor of a perfect fluid. For
this we need to introduce a quantity $u^{\mu}_{(\phi)}$, having properties
similar to the four-velocity of the perfect fluid, that is, satisfying the
conditions $g_{\mu \nu}u^{\mu}_{(\phi)}u^{\nu}_{(\phi)}=1$, and the vector
is timelike. Such a vector can be constructed as \cite{Fa,Ge}
\begin{equation}
u^{\mu}_{(\phi)}=\frac{\nabla ^{\mu}\phi}{\sqrt{\nabla _{\nu}\phi\nabla
^{\nu}\phi}}.
\end{equation}
For such a construction to be physically acceptable, that is, for $%
u^{\mu}_{(\phi)}$ to be real and timelike, the scalar field must satisfy the
conditions that $\nabla ^{\mu}\phi $ is real, and $\nabla _{\nu}\phi\nabla
^{\nu}\phi>0$. Then we can associate to the scalar field an effective energy
density $\rho _{\phi}$ and pressure $p_{\phi}$, given by $\rho
_{\phi}=T^{\mu \nu}_{(\phi)}u_{(\phi) \mu}u_{(\phi) \nu}$, and $%
p_{\phi}=(1/3)\Pi _{\mu}^{\mu}$, where $\Pi_{\mu\nu}=-T_{(\phi) \sigma
\rho}h^{\sigma}_{\mu}h^{\rho}_{\nu}$, where $h^{\sigma}_{\mu}=u^{\sigma}_{(%
\phi)}u_{(\phi)\mu}-\delta _{\mu}^{\sigma}$ is the projection operator.
Then, by using these definitions, one can easily obtain
\begin{equation}
\rho _{\phi}=\frac{1}{2}\nabla _{\mu}\phi \nabla ^{\mu}\phi +U(\phi),
p_{\phi}=\frac{1}{2}\nabla _{\mu}\phi \nabla ^{\mu}\phi -U(\phi).
\end{equation}

The energy - momentum tensor of the scalar field can then be written in a
form similar to the perfect fluid case,
\begin{equation}
T^{\mu \nu}_{(\phi)}=\left( \rho _{\phi}+p_{\phi}\right) u^{\mu
}_{(\phi)}u^{\nu }_{(\phi)}-p_{\phi}g^{\mu \nu}.
\end{equation}

The thermodynamical and physical properties of the mixture of scalar field
and radiation cosmological fluid can be obtained from the total
energy-momentum tensor $T^{\mu \nu }$ of the Universe, given by the sum of
each individual components as
\begin{eqnarray}
T^{\mu \nu }&=&T^{\mu \nu}_{(\phi)}+T^{\mu \nu}_{(rad)}=\left( \rho _{\phi}+p_{\phi}\right) u^{\mu }_{(\phi)}u^{\nu
}_{(\phi)}-p_{\phi}g^{\mu \nu }+  \nonumber \\
&&\left( \rho _{rad}+p_{rad}\right) u^{\mu }_{(rad)}u^{\nu
}_{(rad)}-p_{rad}g^{\mu \nu }.  \label{emtensor}
\end{eqnarray}

The four-velocities of the scalar field and of the radiation fluid are
normalized according to $u^{\mu }_{(\phi)}u_{(\phi)\;\mu }=1$ and $u^{\mu
}_{(rad)}u_{(rad)\;\mu }=1$, respectively. As for the radiation fluid we
will adopt the standard equation of state for the photon gas, $%
p_{rad}=(1/3)\rho _{rad}$.

The total energy-momentum of the mixture of the radiation fluid and scalar field  must satisfy the conservation equation $\nabla _{\mu}T^{\mu \nu}=0$, which gives
\bea
\nabla _{\mu}T^{\mu \nu}&=&\left( \nabla _{\mu}\rho _{\phi}+\nabla _{\mu}p_{\phi}\right) u^{\mu }_{(\phi)}u^{\nu
}_{(\phi)}+\nonumber\\
&&\left( \rho _{\phi}+p_{\phi}\right)\nabla _{\mu}u^{\mu }_{(\phi)}u^{\nu
}_{(\phi)}+\left( \rho _{\phi}+p_{\phi}\right)u^{\mu }_{(\phi)}\nabla _{\mu}u^{\nu
}_{(\phi)}\nonumber\\
&&-\nabla _{\mu}p_{\phi}g^{\mu \nu }+\nabla _{\mu}T^{\mu \nu}_{(rad)}=0.
\eea
After multiplication of the above equation by $u_{(\phi)\nu
}$, and by taking into account the mathematical identity $u_{(\phi)\nu}\nabla _{\mu}u^{(\phi)\nu}=0$, the energy conservation equation of the mixture of two fluids takes the form
\be\label{7n}
\dot{\rho}_{\phi}+3H_{\phi}\left(\rho _{\phi}+p_{\phi}\right)+u_{(\phi)\nu}\nabla _{\mu}T^{\mu \nu}_{(rad)}=0,
\ee
where we have denoted $\dot{(...)} =u_{(\phi)\nu}\nabla ^{\mu}(...)$, while $H_{\phi}=(1/3)\nabla _{\mu}u^{(\phi)\mu}$. Alternatively, by using  the radiation fluid quantities the energy conservation equation can be formulated as
\be\label{8n}
\dot{\rho}_{rad}+3H_{rad}\left(\rho _{rad}+p_{rad}\right)+u_{(rad)\nu}\nabla _{\mu}T^{\mu \nu}_{(\phi)}=0,
\ee
where in the case of radiation  $\dot{(...)} =u_{(rad)\nu}\nabla ^{\mu}(...)$, while $H_{rad}=(1/3)\nabla _{\mu}u^{(rad)\mu}$. Eqs.~(\ref{7n}) and (\ref{8n}) give the evolution of the energy densities of the scalar field and radiation fluid, indicating the possibility of the energy transfer from one component to the other. In order for the two fluids to expand at the same rate the condition $H_{\phi}=H_{rad}$ must be satisfied, giving the condition $\nabla _{\mu}u^{(\phi)\mu}=\nabla _{\mu}u^{(rad)\mu}$. This equation has the obvious solution $u^{(\phi)\mu}=u^{(rad)\mu}$. But if we interpret it mathematically as a partial differential equation for either $u^{(\phi)\mu}$ or $u^{(\phi)\mu}$, then more general solutions with $u^{(\phi)\mu}=u^{(\phi)\mu}\left(u^{(rad)\nu},x^{\lambda}\right)$ are also possible, and their existence is assured by the condition of the existence and unicity of the solutions of first order partial differential equations.

In the standard cosmological models of inflation it is usually assumed that the two cosmological fluids (scalar field and radiation)
are comoving, which implies
\begin{equation}
u_{(\phi)}^{\mu }\equiv u_{(rad)}^{\mu }\equiv u^{\mu }, \mu =0,1,2,3.
\end{equation}
This condition allows to choose for the study of the cosmological dynamics a
single comoving frame, with the result that all the components of the four velocity of all
cosmic constituents can be reformulated as $u^{\mu }=(1,0,0,0)$. In the case
when $u^{\mu }_{(\phi)}= u^{\mu}_{(rad)}\equiv w^{\mu }$, the
total energy-momentum tensor of the scalar field plus radiation system takes the
simple form
\begin{equation}
T^{\mu \nu }=\left( \rho _{\phi}+p_{\phi}+\rho _{rad}+p_{rad}\right) w^{\mu
}w^{\nu }-\left(p_{\phi}+p_{rad}\right)g^{\mu \nu}.
\end{equation}

In this case the scalar field-radiation fluid system reduces to an isotropic
effective single fluid system. Therefore, if the scalar field and the photon gas
have identical four-velocities, the thermodynamic parameters of the scalar
field-radiation two fluid system are obtained through a simple addition of pressures and
enthalpies of the two individual components. In this case one can always introduce a comoving frame
to describe the properties of the physical system. In the comoving frame the
components of the four-velocity are $w^{\mu }=(1,0,0,0)$, with the
components of the total energy-momentum tensor of the two fluids obtained as $%
T_0^0= \left( \rho _{\phi}+\rho _{rad}\right)\delta _0^0$, and $%
T_i^i=-\left(p_{\phi}+p_{rad}\right)\delta _i^i$, $i=1,2,3$, respectively, with no summation
upon $i$.

However, in the present study of warm inflation we will abandon the
condition of the comoving motion, by assuming that during at least some intervals of the
cosmological expansion, the scalar field and the radiation components of the
warm inflationary Universe could have had different four-velocities, and hence
\begin{equation}
u_{(\phi)}^{\mu }\neq u_{(rad)}^{\mu },\; \mu =0,1,2,3.
\end{equation}

In such a situation it is not possible to introduce a comoving frame, so
that the thermodynamical quantities are constructed as the sum of the
thermodynamical parameters of each component of the mixture. In our analysis
of the warm inflationary scenarios in noncomoving frames we  assume that
the energy density and the pressure of the scalar field satisfy the
condition $\rho _{\phi}+p_{\phi}\geq 0$, that is, the scalar field cannot be
interpreted as a simple cosmological constant.

\subsection{Single anisotropic fluid representation of two fluid systems}

The analysis of warm inflationary cosmological models in which the
total matter content is described by an energy-momentum tensor having the
representation given by Eq.~(\ref{emtensor}) can be simplified essentially
if we represent it in the standard form of the energy-momentum tensor of
single perfect anisotropic fluids. This can be achieved by using  the
four-velocity transformations
\be
u_{(\phi )}^{\mu }\rightarrow u_{(\phi
)}^{\ast \mu }, u_{(rad)}^{\mu }\rightarrow u_{(rad)}^{\ast \mu },
\ee
respectively, with the transformation matrix given by \cite{Le,Le1,Ol, He}
\begin{eqnarray}\label{tel}
\left(
\begin{array}{c}
u_{(\phi )}^{\ast \mu } \\
u_{(rad)}^{\ast \mu }%
\end{array}%
\right) &=&\left(
\begin{array}{cc}
\cos \alpha  & \sqrt{\frac{\rho _{rad}+p_{rad}}{\rho _{\phi }+p_{\phi }}}%
\sin \alpha  \\
-\sqrt{\frac{\rho _{\phi }+p_{\phi }}{\rho _{rad}+p_{rad}}}\sin \alpha  &
\cos \alpha
\end{array}%
\right) \times \nonumber\\
&&\left(
\begin{array}{c}
u_{(\phi )}^{\mu } \\
u_{(rad)}^{\mu }%
\end{array}%
\right) ,
\end{eqnarray}%
or, equivalently,
\bea\label{te11}
\left(
\begin{array}{c}
u_{(\phi )}^{\ast \mu } \\
u_{(rad)}^{\ast \mu }%
\end{array}%
\right) &=&\left(
\begin{array}{cc}
\cos \alpha  & \sqrt{\frac{4}{3}\frac{\rho _{rad}}{\nabla _{\mu }\phi \nabla
^{\mu }\phi }}\sin \alpha  \\
-\sqrt{\frac{3}{4}\frac{\nabla _{\mu }\phi \nabla ^{\mu }\phi }{\rho _{rad}}}%
\sin \alpha  & \cos \alpha
\end{array}%
\right) \nonumber\\
&&\times \left(
\begin{array}{c}
u_{(\phi )}^{\mu } \\
u_{(rad)}^{\mu }%
\end{array}%
\right).
\eea

The transformation introduced above represents a nonsingular rotation of the four-vector velocities of the two components in
the $\left( u_{(\phi )}^{\mu },u_{(rad)}^{\mu }\right) $ velocity space.
Explicitly, the transformations (\ref{te11}) take the form
\begin{equation}
u_{(\phi )}^{\mu }\rightarrow u_{(\phi )}^{\ast \mu }=u_{(\phi )}^{\mu }\cos
\alpha +\sqrt{\frac{4}{3}\frac{\rho _{rad}}{\nabla _{\mu }\phi \nabla ^{\mu
}\phi }}u_{(rad)}^{\mu }\sin \alpha ,  \label{te1}
\end{equation}%
\begin{equation}
u_{(rad)}^{\mu }\rightarrow u_{(rad)}^{\ast \mu }=-\sqrt{\frac{3}{4}\frac{%
\nabla _{\mu }\phi \nabla ^{\mu }\phi }{\rho _{(rad)}}}u_{(\phi )}^{\mu
}\sin \alpha +u_{(rad)}^{\mu }\cos \alpha \,,  \label{te2}
\end{equation}%
respectively.

The transformations in the velocity space given by Eq.~(\ref{tel}) leave the
quadratic form $\left( \rho _{\phi }+p_{\phi }\right) u_{(\phi )}^{\mu
}u_{(\phi )}^{\nu }+\left( \rho _{rad}+p_{rad}\right) u_{(rad)}^{\mu
}u_{(rad)}^{\nu }$ invariant. Thus,
\begin{equation}
T^{\mu \nu }\left( u_{(\phi )}^{\mu },u_{(rad)}^{\nu }\right) =T^{\mu \nu
}\left( u_{(\phi )}^{\ast \mu },u_{(rad)}^{\ast \nu }\right) .
\end{equation}%

As a next step in our analysis we choose $u_{(\phi )}^{\ast \mu }$ and $%
u_{(rad)}^{\ast \mu }$ so that one becomes timelike, while the other one is
spacelike. Moreover, we also assume that the two new four-vector velocities
satisfy the orthogonality condition
\begin{equation}
u_{(\phi )}^{\ast \mu }u_{(rad)\;\mu }^{\ast }=0.  \label{cond}
\end{equation}

By using Eqs.~(\ref{te1})-(\ref{te2}) as well as the orthogonality condition
as given by Eq.~(\ref{cond}), we find for the angle of rotation the
expression
\bea\label{alpha}
\tan 2\alpha &=&2\frac{\sqrt{\left(\rho _{\phi}+p_{\phi}\right)\left(\rho_{rad}+p_{rad}\right)}}{\rho_{\phi}+p_{\phi}-\rho_{rad}-p_{rad}}u_{(\phi
)}^{\mu }u_{(rad)\;\mu }=\nonumber\\
&&2\frac{\sqrt{4\left(\nabla _{\mu }\phi \nabla ^{\mu }\phi \right)\rho
_{rad}/3}}{\nabla _{\mu }\phi \nabla ^{\mu }\phi -4\rho _{rad}/3}u_{(\phi
)}^{\mu }u_{(rad)\;\mu }.
\eea

If the angle $\alpha $ has a different form, and it is not given by Eq.~(\ref%
{alpha}), we cannot find a transformed spacelike $u_{(\phi )}^{\ast \mu } $,
and a timelike $u_{(rad)}^{\ast \mu }$ scalar field and radiation fluid
velocity, respectively. Note that if $\rho _{\phi }+p_{\phi }=0$, a case
that corresponds to the presence of a cosmological constant, the rotation angle becomes $%
\alpha =0$. Consequently, in the velocity space, the rotation reduces to the
identical transformation. Hence, if a cosmological constant does exist,
since $\rho _{\phi }+p_{\phi }=0$, the term $\left( \rho _{\phi }+p_{\phi
}\right) u_{(\phi )}^{\mu }u_{(\phi )}^{\nu }$ in the energy-momentum tensor
of the scalar field is identically equal to zero, $\left( \rho _{\phi
}+p_{\phi }\right) u_{(\phi )}^{\mu }u_{(\phi )}^{\nu }\equiv 0$. Therefore,
the four-velocity of the cosmological constant is not present in the above
introduced formalism. Hence the effective energy - momentum tensor of a
cosmological system consisting of a simple cosmological constant $\Lambda={\rm constant}$ plus a
radiation fluid can be equivalently described by the energy-momentum tensor
of a single isotropic fluid.

Next, we introduce the new set of quantities $\left(V^{\mu }, \chi ^{\mu }, \varepsilon, \Psi, \Pi\right)$ defined according to
\begin{equation}
V^{\mu }=\frac{u_{(\phi )}^{\ast \mu }}{\sqrt{u_{(\phi )}^{\ast \alpha
}u_{(\phi )\;\alpha }^{\ast }}},\chi ^{\mu }=\frac{u_{(rad)}^{\ast \mu }}{%
\sqrt{-u_{(rad)}^{\ast \alpha }u_{(rad)\;\alpha }^{\ast }}},
\end{equation}%
\bea
\hspace{-1.1cm}&&\varepsilon =T^{\mu \nu }V_{\mu }V_{\nu }=\left( \rho _{\phi }+p_{\phi
}\right) u_{(\phi )}^{\ast \alpha }u_{(\phi )\;\alpha }^{\ast }-\left(
p_{\phi }+p_{rad}\right) ,
\eea
\bea
\hspace{-1.2cm}&&\Psi =T^{\mu \nu }\chi _{\mu }\chi _{\nu }=\left( p_{\phi}+p_{rad}\right)
-
\left( \rho _{rad}+p_{rad}\right) u_{(rad)}^{\ast \alpha }u_{(rad)\;\alpha
}^{\ast },
\eea
and
\begin{equation}
\hspace{-6.0cm}\Pi =p_{\phi }+p_{rad}\;,
\end{equation}%
respectively, where from a physical point of view $\varepsilon $ can be interpreted as the energy density
and $\Psi $ as the radial pressure of an anisotropic Universe. Then, by adopting this interpretation,  it turns
out that the energy-momentum tensor of the two-component system consisting of a noncomoving scalar field and
a radiation fluid is obtained in the form
\begin{equation}
T^{\mu \nu }=\left( \varepsilon +\Pi \right) V^{\mu }V^{\nu }-\Pi g^{\mu \nu
}+\left( \Psi -\Pi \right) \chi ^{\mu }\chi ^{\nu },  \label{tens}
\end{equation}%
where $V^{\mu }V_{\mu }=1=-\chi ^{\mu }\chi _{\mu }$ and $\chi ^{\mu }V_{\mu
}=0$ \cite{Le,Le1,Ol, He}.

Hence, the total energy-momentum tensor of the two-fluid warm
inflationary cosmological model, in which the two components have different
four-velocities, given by Eq.~(\ref{tens}), has acquired the standard form of the
energy-momentum tensor for perfect anisotropic fluids, which have already been extensively investigated in the literature \cite{He}.

Moreover, the energy density $\varepsilon $ and the radial pressure $\Psi $ of the
scalar field and radiation  fluid filled inflationary Universe can be obtained as
\begin{eqnarray}\label{eps1}
\hspace{-0.3cm}&&\varepsilon =\frac{1}{2}\left( \rho _{\phi }-p_{\phi }+\rho
_{rad}-p_{rad}\right) + \frac{1}{2}\times  \nonumber  \label{d1} \\
\hspace{-0.3cm}&&\Bigg\{\left( \rho _{\phi }+p_{\phi }+\rho _{rad}+p_{rad}\right)
^{2}+4\left( \rho _{\phi }+p_{\phi }\right) \left( \rho
_{rad}+p_{rad}\right) \nonumber\\
\hspace{-0.3cm}&&\times \left[ \left( u_{(\phi )}^{\mu }u_{(rad)\;\mu }\right)
^{2}-1\right] \Bigg\}^{1/2},
\end{eqnarray}%
and
\begin{eqnarray}\label{psin}
\hspace{-0.3cm}&&\Psi =-\frac{1}{2}\left( \rho _{\phi }-p_{\phi }+\rho
_{rad}-p_{rad}\right) + \frac{1}{2} \times \nonumber   \\
\hspace{-0.3cm}&&\Bigg\{\left( \rho _{\phi }+p_{\phi }+\rho _{rad}+p_{rad}\right)
^{2}+4\left( \rho _{\phi }+p_{\phi }\right) \left( \rho
_{rad}+p_{rad}\right) \nonumber\\
\hspace{-0.3cm}&&\times \left[ \left( u_{(\phi )}^{\mu }u_{(rad)\;\mu }\right)
^{2}-1\right] \Bigg\}^{1/2},
\end{eqnarray}%
respectively \cite{Le,Le1,Ol, He}.  Explicitly, the energy density and the
radial pressure of the noncomoving scalar field - radiation fluid system are given by
\begin{eqnarray}
\hspace{-0.5cm}\varepsilon &=&U(\phi )+\frac{\rho _{rad}}{3}+  \frac{1}{2}\Bigg\{\left( \nabla _{\mu }\phi \nabla ^{\mu }\phi +\frac{4}{3}%
\rho _{rad}\right) ^{2}+\nonumber\\
\hspace{-0.5cm}&&\frac{16}{3}\rho _{rad}\nabla _{\mu }\phi \nabla ^{\mu }\phi
\left[ \left( u_{(\phi )}^{\mu }u_{(rad)\;\mu }\right) ^{2}-1%
\right] \Bigg\}^{1/2},
\end{eqnarray}%
and
\begin{eqnarray}
\Psi &=&-U(\phi )-\frac{\rho _{rad}}{3}+  \frac{1}{2}\Bigg\{\left( \nabla _{\mu }\phi \nabla ^{\mu }\phi +\frac{4}{3}%
\rho _{rad}\right) ^{2}+\nonumber\\
&&\frac{16}{3}\rho _{rad}\nabla _{\mu }\phi \nabla ^{\mu }\phi
\left[ \left( u_{(\phi )}^{\mu }u_{(rad)\;\mu }\right) ^{2}-1%
\right] \Bigg\}^{1/2},
\end{eqnarray}
respectively.

The energy density of the anisotropic effective fluid, describing  the noncomoving mixture of the scalar
field and of the radiation fluid, given by Eq.~(\ref{d1}), depends
explicitly on the expressions of the four-velocities of the two fluids.
Therefore, it follows that the energy density of the rotated fluids is also a function of both the
kinetic energy of the scalar field and of the radiation fluid, respectively. Such a
functional dependence implies that the energy in the rotated four-velocity
space is not equal to the sum of the energy densities of the scalar
field and of radiation, respectively. When $u_{(\phi )}^{\mu
}=u_{(rad)}^{\mu }$, then it follows that $u_{(\phi )}^{\mu }u_{(rad)\;\mu }=1$, and therefore the
expressions of the effective energy and pressure of the effective fluid
can be found as the direct sum of the energy densities of the scalar field and of the
radiation fluid, respectively, thus obtaining $\varepsilon =\rho _{\phi }+\rho
_{rad}$, $\Psi =p_{\phi }+p_{rad}$. Moreover, in this case in order to describe cosmological dynamics one can adopt a frame
comoving with both components.

Since the two components of the inflationary Universe, the scalar field and radiation, have
distinct four-velocities, we can write
\begin{equation}\label{bdef}
u_{(\phi )}^{\mu }u_{(rad)\;\mu }=1+\frac{b}{2},
\end{equation}%
where generally $b$ is an arbitrary function of the thermodynamic parameters (energy densities and pressures) of the scalar field and of the radiation, respectively.
The exact expression of $b=b\left(\rho _{\phi},p_{\phi},\rho _{rad},p_{rad}\right)$ can be found from Eq.~(%
\ref{alpha}), from which we obtain first
\begin{equation}
u_{(\phi )}^{\mu }u_{(rad)\;\mu }=\frac{1}{2}\Bigg[\sqrt{\frac{%
\rho _{\phi }+p_{\phi }}{\rho _{rad}+p_{rad}}}-\sqrt{\frac{\rho
_{rad}+p_{rad}}{\rho _{\phi }+p_{\phi }}}\Bigg]\tan 2\alpha .
\end{equation}%
Then the dependence of $b$ on the thermodynamic parameters can be determined immediately as
\begin{equation}\label{27}
b =\left[ \sqrt{\frac{\rho _{\phi }+p_{\phi }}{\rho _{rad}+p_{rad}}}-\sqrt{%
\frac{\rho _{rad}+p_{rad}}{\rho _{\phi }+p_{\phi }}}\right] \tan 2\alpha -2.
\end{equation}

By using the explicit expression of $b$, Eqs.~(\ref{d1}) and (\ref{psin}) can be reformulated in the form
\begin{eqnarray} \label{epsn}
\hspace{-0.6cm}\varepsilon &=&\frac{1}{2}\left( \rho _{\phi }-p_{\phi }+\rho
_{rad}-p_{rad}\right) +  \nonumber  \\
\hspace{-0.6cm} &&\frac{1}{2}\left( \rho _{\phi }+p_{\phi }+\rho
_{rad}+p_{rad}\right) \times  \nonumber \\
\hspace{-0.6cm} &&\Bigg[1+4b\left(1+\frac{b}{4}\right) \frac{\left( \rho
_{\phi }+p_{\phi }\right) \left( \rho _{rad}+p_{rad}\right) }{\left( \rho
_{\phi }+p_{\phi }+\rho _{rad}+p_{rad}\right) ^{2}}\Bigg]^{1/2},
\end{eqnarray}%
and
\begin{eqnarray}\label{psin1}
\hspace{-0.7cm}\Psi &=&-\frac{1}{2}\left( \rho _{\phi }-p_{\phi }+\rho
_{rad}-p_{rad}\right) + \nonumber \\
\hspace{-0.7cm} &&\frac{1}{2}\left( \rho _{\phi }+p_{\phi }+\rho
_{rad}+p_{rad}\right) \times  \nonumber \\
\hspace{-0.7cm} &&\Bigg[1+4b\left( 1+\frac{b}{4}\right) \frac{\left( \rho
_{\phi }+p_{\phi }\right) \left( \rho _{rad}+p_{rad}\right) }{\left( \rho
_{\phi }+p_{\phi }+\rho _{rad}+p_{rad}\right) ^{2}}\Bigg]^{1/2},
\end{eqnarray}%
respectively. Explicitly, we obtain
\begin{eqnarray}
\hspace{-0.6cm}\varepsilon &=&U(\phi )+\frac{\rho _{rad}}{3} + \frac{1}{2}\left( \nabla _{\mu }\phi \nabla ^{\mu }\phi +\frac{4}{3}\rho
_{rad}\right) \times  \nonumber \\
\hspace{-0.6cm} &&\sqrt{1+\frac{16b}{3}\left( 1+\frac{b}{4}\right) \frac{%
\nabla _{\mu }\phi \nabla ^{\mu }\phi \rho _{rad}}{\left( \nabla _{\mu }\phi
\nabla ^{\mu }\phi +\frac{4}{3}\rho _{rad}\right) ^{2}}},
\end{eqnarray}%
and
\begin{eqnarray}
\hspace{-0.6cm}\Psi &=&-U(\phi )-\frac{\rho _{rad}}{3}+  \frac{1}{2}\left( \nabla _{\mu }\phi \nabla ^{\mu }\phi +%
\frac{4}{3}\rho _{rad}\right) \times  \nonumber \\
\hspace{-0.6cm} &&\sqrt{1+\frac{16b}{3}\left( 1+\frac{b}{4}\right) \frac{%
\nabla _{\mu }\phi \nabla ^{\mu }\phi \rho _{rad}}{\left( \nabla _{\mu }\phi
\nabla ^{\mu }\phi +\frac{4}{3}\rho _{rad}\right) ^{2}}}.
\end{eqnarray}

Next we adopt the  assumption that the physical parameters of the scalar field and of the
radiation fluid satisfy the condition
\begin{equation}\label{32}
4b\left( 1+\frac{b}{4}\right) \frac{\left( \rho _{\phi }+p_{\phi }\right)
\left( \rho _{rad}+p_{rad}\right) }{\left( \rho _{\phi }+p_{\phi }+\rho
_{rad}+p_{rad}\right) ^{2}}\ll 1,
\end{equation}%
or%
\begin{equation}
b\left( 1+\frac{b}{4}\right) \frac{\nabla _{\mu }\phi \nabla ^{\mu }\phi
\rho _{rad}}{\left( \nabla _{\mu }\phi \nabla ^{\mu }\phi +\frac{4}{3}\rho
_{rad}\right) ^{2}}\ll \frac{3}{16}.
\end{equation}%
Then, after series expanding the square root in Eqs.~(\ref{epsn}) and (\ref{psin}%
), and keeping only the first order of approximation, it follows that the
energy density, the radial and the tangential pressures of the inflationary Universe
filled with a scalar field and a radiation fluid can be obtained as
\begin{eqnarray}\label{epsa}
\varepsilon  &=&\rho _{\phi }+\rho _{rad}+F\left(\rho_{\phi},\rho _{rad}\right)=  \nonumber\\
&&\frac{1}{2}\nabla _{\mu }\phi \nabla ^{\mu }\phi +U(\phi )+\rho _{rad}+F\left(\rho_{\phi},\rho _{rad}\right),
\end{eqnarray}%
\begin{eqnarray}
\Psi  &=&p_{\phi }+p_{rad}+F\left(\rho _{\phi},\rho _{rad}\right)= \nonumber\\
&&\frac{1}{2}\nabla _{\mu }\phi \nabla ^{\mu }\phi -U(\phi )+p_{rad}+F\left(\rho_{\phi},\rho _{rad}\right),
\end{eqnarray}
and
\begin{equation}
\Pi =p_{\phi }+p_{rad},
\end{equation}%
respectively, where we have denoted
\bea
F\left(\rho _{\phi},\rho _{rad}\right)&=&2b\left( 1+\frac{b}{4}\right)
\frac{\left( \rho _{\phi }+p_{\phi }\right) \left( \rho
_{rad}+p_{rad}\right) }{\left( \rho _{\phi }+p_{\phi }+\rho
_{rad}+p_{rad}\right) }=\nonumber\\
&&\frac{8b}{3}\left( 1+\frac{b}{4}\right) \frac{\nabla _{\mu }\phi \nabla
^{\mu }\phi \rho _{rad}}{\left( \nabla _{\mu }\phi \nabla ^{\mu }\phi +4\rho
_{rad}/3\right) }.
\eea

By assuming that $\rho _{\phi}+p_{\phi}>>\rho _{rad}+p_{rad}$, we obtain for $F$ the expression
\be
F\left(\rho _{\phi},\rho _{rad}\right)\approx \frac{8b}{3}\left( 1+\frac{b}{4}\right)\rho _{rad}.
\ee
I the limit $b/4>>1$, we obtain
\be
F\left(\rho _{\phi},\rho _{rad}\right)\approx \frac{2}{3}b^2\rho _{rad},
\ee
while in the opposite limit $b/4<<1$, $F\left(\rho _{\phi},\rho _{rad}\right)$ can be approximated as
\be
F\left(\rho _{\phi},\rho _{rad}\right)\approx \frac{8b}{3}\rho _{rad}.
\ee

 As one can observe easily from Eq.~(\ref{epsa}), the total energy
density of the non-comoving scalar field and radiation fluid filled inflationary Universe
is different from the simple sum of the energy densities of the two cosmological fluids.
Moreover, a coupling between the energy densities of the radiation and of
the scalar field is naturally generated in this scenario.

\section{Warm inflation with noncommoving scalar field and radiation}\label{sect3}

In the present Section, we will investigate in detail the cosmological properties of the
non-comoving scalar field - radiation fluid physical system. More exactly,
we will consider this model in the framework of the warm inflationary
scenario, in which it is assumed that the very early Universe consisted of a
scalar field that decayed into a radiation fluid. We will adopt the fundamental assumption  that the
four-velocity of the scalar field was not exactly equal with the
four-velocity of the photon gas. Therefore, in such a situation, one cannot
introduce a comoving frame to describe the global evolution. Moreover, the
noncomoving two fluid system becomes anisotropic, and the total energy and
pressure of the system contains an interaction term between the energy
densities and pressures of the scalar field and the radiation fluid.

As a first step in our analysis of the noncomoving warm inflationary scenario, we
obtain the gravitational field equations describing  the
anisotropic evolution of the early Universe, filled with two interacting scalar field
and radiation fluids. We analyze then the general properties of the theoretical model,
and we show that the non-comoving nature of the cosmological dynamics
induces some significant anisotropic effects in the evolution of both shear and
and expansion cosmological parameters. We also consider specific noncomoving warm
inflationary models, and the general solution of the gravitational field equations,
describing the scalar field-radiation fluid mixture, is obtained.

\subsection{Brief review of the warm inflationary scenario}

 The warm inflationary model \cite{w1,w2} is an interesting and elegant theoretical approach, representing a powerful alternative to the standard inflation and reheating scenarios. Analogously to cold inflationary models, in warm inflation the Universe also evolves through an accelerated, de Sitter type, very early expansionary phase. The de Sitter stage is triggered due to the presence of a scalar field,  which represents the major component of the early Universe. But, in opposition to the cold inflationary scenario, in addition to the primordial scalar field, a matter component,  assumed to exist in the form of a photon fluid, is also present. The matter component is created by the scalar field all along the accelerating expansion phase. An essential assumption of the warm inflationary scenario is that during the very early cosmological evolution, these two different components interact through a dynamical process. We assume that in the early Universe the cosmological evolution is still governed by the standard Friedmann equations, given by
 \be\label{w01}
 3H^2=\frac{1}{M_P^2}\left(\rho _{\phi }+\rho _{rad}\right),
 \ee
 \be\label{w02}
 2\dot{H}=-\frac{1}{M_P^2}\left(\dot{\phi}^2+\frac{4}{3}\rho_{rad}\right),
 \ee
where by $M_P$ we have denoted the Planck mass. As usual, the energy density and the pressure of the scalar field are given,  by $\rho _{\phi}=\dot{\phi}^2/2+U(\phi)$ and $p _{\phi}=\dot{\phi}^2/2-U(\phi)$, respectively, where by $U(\phi)$ we have denoted the scalar field self-interaction potential.  As a result of the dynamic decay of the scalar field, which, from thermodynamic point of view, is essentially a dissipative mechanism, an energy transfer process from the field to radiation takes place. The decay/creation processes are described by the following energy balance equations,
\be\label{w02a}
\dot{\rho}_{\phi}+3H\left(\rho _{\phi}+p_{\phi}\right)=-\Gamma \dot{\phi}^2,
\ee
\be\label{w03}
\dot{\rho}_{rad}+3H\left(\rho _{rad}+p_{rad}\right)=\Gamma \dot{\phi}^2,
\ee
where by $\Gamma$  we have denoted the dissipation coefficient. With the use of the explicit expressions of the energy density and pressure of the scalar field, Eq.~(\ref{w02a}) becomes the generalized Klein-Gordon evolution equation for the scalar field,
\be\label{KG0}
\ddot{\phi}+3H\left(1+Q\right)\dot{\phi}+U'(\phi)=0,
\ee
where we have denoted $Q=\Gamma /3H$. In order to simplify the mathematical formalism we assume first that the  expansion of the Universe is quasi-de Sitter. Secondly, we assume that the energy density of the scalar field is much bigger than the energy density of the radiation, and hence that the condition $\rho _{\phi}>>\rho _{rad}$ holds. Moreover, we assume that the potential energy term of the energy density of the scalar field dominates the kinetic one, and hence $\rho _{\phi}\approx U(\phi)$. Therefore,  Eqs.~(\ref{w01}), (\ref{w03}), and (\ref{KG0}) can be approximated as
\be
3H^2\approx \frac{1}{M_P^2}U(\phi), \dot{\phi}\approx -\frac{U'(\phi)}{3H(1+Q)},
\ee
\be\label{w04}
\rho _{rad}=C_\gamma T^4\approx \frac{\Gamma}{4H}\dot{\phi}^2,
\ee
where $C_\gamma=\pi^2 g_\star / 30$ is the Stefen-Boltzman constant, while $g_\star$ denotes the number of degrees of freedom of the photon field. In order to obtain Eq.~(\ref{w04}) we have also used two other approximations, $\dot{\rho}_{rad}<<H\rho_{rad}$, and $\dot{\rho}_{rad}<<\Gamma \dot{\phi}^2$, respectively.

To describe the accelerating/decelerating evolution of the Universe,  we introduce the deceleration parameter $q$, defined according to the expression
\be
q=\frac{d}{dt}\frac{1}{H}-1.
\ee
If $q<0$, the expansion is accelerating, while $q>0$ indicates a decelerating cosmological evolution. By using Eqs.~(\ref{w01}) and (\ref{w02}), we  obtain the deceleration parameter of the standard warm inflation theory in the form
\be
q=\frac{1}{2}\left[1+\frac{3\left(p_{\phi}+p_{rad}\right)}{\rho _{\phi}+\rho_{rad}}\right].
\ee

Similar information to the one provided by the deceleration parameter $q$ can be extracted from the quantity  $\epsilon_{ H} \equiv d \ln H/d\mathcal{N} $, where $\mathcal{N}$ denotes the number of e-folds.  Note that $\epsilon_{ H}$ is related to the important cosmological parameter $\epsilon$, which provides important information on the validity and properties of the slow-roll approximation in inflationary models.

\subsection{The gravitational field equations in the presence of non-comoving scalar field and radiation fluid}

In the case of a noncomoving scalar field-radiation two fluid mixture,  the components of the energy-momentum tensor are given by  Eqs.~(\ref{tens}). Hence, for such a physical configuration,  the Einstein gravitational field equations can be written in the form \cite{Ol}
\begin{equation}
R_{\mu \nu }V^{\mu }V^{\nu }=\frac{1}{2}\left( \varepsilon +2\Pi +\Psi
\right) ,
\end{equation}%
\begin{equation}
R_{\mu \nu }V^{\mu }h_{\sigma }^{\nu }=0,
\end{equation}%
\begin{eqnarray}
R_{\mu \nu }h_{\sigma }^{\mu }h_{\lambda }^{\nu } &=&\frac{1}{2}\left[
\varepsilon -\frac{1}{3}\left( 2\Pi +\Psi \right) \right] h_{\sigma \lambda
}+  \nonumber \\
&&\left( \Psi -\Pi \right) \left( \chi _{\sigma }\chi _{\lambda }-\frac{1}{3}%
h_{\sigma \lambda }\right) ,
\end{eqnarray}%
where $h^{\mu \nu }=g^{\mu \nu }-V^{\mu }V^{\nu }$, and $R_{\mu \nu}$ is the Ricci tensor. Moreover, the conservation of the total
energy-momentum tensor of the matter, $\nabla _{\mu }T_{\nu }^{\mu }=0$, yields the equations \cite{Ol}
\begin{equation}
\dot{\varepsilon}+\left( \varepsilon +\Pi \right) \nabla _{\mu }V^{\mu
}-\left( \Psi -\Pi \right) \chi ^{\prime \mu }V_{\mu }=0,
\end{equation}%
and
\begin{eqnarray}
&&\left( \varepsilon +\Pi \right) \dot{V}^{\mu }+h^{\mu \nu }\nabla _{\nu
}\Pi +\left( \Psi -\Pi \right) ^{\prime }\chi ^{\mu }+  \nonumber \\
&&\left( \Psi -\Pi \right) \left( \nabla _{\nu }\chi ^{\nu }\chi ^{\mu
}+\chi _{\nu }^{\prime }h^{\nu \mu }\right) =0,
\end{eqnarray}%
respectively, where we have defined the overdot and the prime according to $\dot{\left( \;\right) }%
=V^{\mu }\nabla _{\mu }\left( \;\right) $, and $\left( \;\right) ^{\prime
}=\chi ^{\mu }\nabla _{\mu }\left( \;\right) $, respectively.

\subsubsection{The field equations for a Bianchi type I geometry}

In the following, we adopt the simplifying assumption of the large scale homogeneity of
the very early inflationary Universe. Therefore all the cosmological
quantities (four-velocities, energy densities, pressures, etc. ) can be
only functions of the universal cosmological time $t$. The assumption of the homogeneity of
the early Universe permits the introduction of a frame that is "comoving"
with the auxiliary quantities $V$ and $\chi $. In the Cartesian coordinate
system with coordinates  $x^{0}=t$, $x^{1}=x$, $x^{2}=y$, and $x^{3}=z$, respectively, we may rescale
all the components of the four velocities $V^{\mu }$ and $\chi ^{\mu \text{ }}$
according to  $V^{1}=V^{2}=V^{3}=0$, $V^{0}V_{0}=1$, and $\chi ^{0}=\chi ^{1}=\chi
^{2}=0$, $\chi ^{3}\chi _{3}=-1$, respectively \cite{Le, Ol, He}. Therefore,
in the frame comoving with $V^{\mu }$ and $\chi ^{\mu }$, the
components of the energy-momentum tensor of the non-comoving scalar field
and radiation fluids take the simple form,
\begin{equation}
T_{0}^{0}=\varepsilon ,\qquad T_{1}^{1}=T_{2}^{2}=-\Pi ,\qquad
T_{3}^{3}=-\Psi ,  \label{tenss}
\end{equation}%
corresponding to a standard anisotropic fluid. In the relations above $\varepsilon $ represents the total energy-density of the mixture of fluids, $%
\Pi =P_{x}=P_{y}$ is the thermodynamic pressure along the $x$ and $y$ directions, while $%
\Psi =P_{z}$ is the total cosmological pressure along the $z$ direction.
Since the energy-momentum tensor of the noncomoving two-fluid scalar field - photon gas cosmological
mixture corresponds to an anisotropic fluid, it turns out that at the cosmological level the
spacetime geometry must also be anisotropic (but homogeneous). The components  of the
energy-momentum tensor, given by Eqs.~(\ref{tenss}), show that along the $z-
$ direction the total pressure $\Psi $ is generally different as compared
with the pressure $\Pi $ exerted on the $x-y$ plane.

 The simplest geometry displaying the symmetries of the
energy-momentum tensor in a homogeneous Universe containing a scalar field - radiation fluid
mixture, is the spatially flat Bianchi type I geometry, with
the line element written as
\begin{equation}
ds^{2}=dt^{2}-a_{1}^{2}(t)dx^{2}-a_{2}^{2}(t)dy^{2}-a_{3}^{2}(t)dz^{2},
\label{7}
\end{equation}%
where $a_{i}(t)$ ($i=1,2,3$) are the time dependent only directional scale factors, describing the anisotropic expansion of the Universe. In the
following we introduce the notations
\begin{equation}\label{55a}
V=a_{1}a_{2}a_{3},\qquad H_{i}=\frac{\dot{a}_{i}}{a_{i}},\qquad i=1,2,3
\end{equation}%
and
\begin{equation}\label{56a}
H=\frac{1}{3}\left( \sum_{i=1}^{3}H_{i}\right) =\frac{\dot{V}}{3V}
\end{equation}%
respectively. An important observational parameter, the expansion parameter $%
\theta $, is defined as $\theta =3H$.

In the Bianchi type I geometry, with line element given by Eq. (\ref%
{7}), the gravitational field equations for the noncomoving scalar field -
photon gas mixture are given by
\begin{equation}
3\dot{H}+H_{1}^{2}+H_{2}^{2}+H_{3}^{2}=-\frac{1}{2}\left( \varepsilon +\Psi
+2\Pi \right) ,  \label{8}
\end{equation}%
\begin{equation}
\frac{1}{V}\frac{d}{dt}\left( VH_{i}\right) =\frac{1}{2}\left( \varepsilon
-\Pi \right) ,\qquad i=1,2,  \label{9}
\end{equation}%
and
\begin{equation}
\frac{1}{V}\frac{d}{dt}\left( VH_{3}\right) =\frac{1}{2}\left( \varepsilon
-\Psi \right) ,  \label{1011}
\end{equation}%
respectively. The evolution equation
of the energy density of the noncomoving warm inflationary model follows from the conservation of the total energy-momentum tensor of
the scalar field - radiation fluid system, and it can be written as
\begin{equation}\label{epst}
\dot{\varepsilon}+3\left( \varepsilon +\Pi \right) H-\left( \Psi -\Pi
\right) H_{3}=0.
\end{equation}

By adding Eqs.~(\ref{9}) and (\ref{1011}) we immediately obtain the evolution equation of the Hubble function,
\begin{equation}
\dot{H}+3H^{2}=\frac{1}{2}\varepsilon -\frac{1}{3}\left(\frac{\Psi}{2}+\Pi
\right),  \label{10}
\end{equation}%
which can be reformulated in an equivalent way as
\begin{equation}\label{41}
\ddot{V}+\left(\Pi+\frac{1}{2}\Psi -\frac{3}{2}\varepsilon \right)V=0.
\end{equation}

\subsection{Warm inflation with noncomoving cosmological fluids}

By using the approximate expression (\ref{epsa}) for the total energy density of the noncomoving scalar field and radiation, we obtain the conservation equation in the cosmologically expanding anisotropic scalar field-radiation mixture as
\bea\label{epstw}
\hspace{-0.4cm}&&\dot{\rho}_{\phi}+\dot{\rho}_{rad}+3H\left(\rho_{\phi}+p_{\phi}+\rho _{rad}+p_{rad}+2F\right)+\nonumber\\
\hspace{-0.4cm}&&\dot{F}-FH_3=0.
\eea

We assume now, similarly to the standard warm inflationary model,  that the scalar field decays into radiation. Accordingly, we can reformulate Eq.~(\ref{epstw}) so that it describes the energy and matter transfer from scalar field to radiation as two separate balance equations,
\be\label{61}
\dot{\rho}_{\phi}+3H\left(\rho _{\phi}+p_{\phi}\right)=-2\dot{F}-12FH=-\Gamma _{\phi},
\ee
and
\be\label{62}
\dot{\rho}_{rad}+3H\left(\rho_{rad}+p_{rad}\right)=\dot{F}+6FH+FH_3=\Gamma _{rad},
\ee
where $\Gamma _{\phi}$ and $\Gamma _{rad}$ are the scalar field and radiation decay and creation rates, respectively. To build specific cosmological models we need to estimate the expression of $F\left(\rho _{\phi},\rho _{rad}\right)$. In the following we will assume that the parameter $b$, describing, according to Eq.~(\ref{bdef}), the differences in the four-velocities of the photon fluid and the scalar field is large, that is, $b>>1$. Consequently, at least at the initial phases of the cosmological expansion, there were important differences in the four-velocities of the two components of the fluid mixture.

By assuming that $\rho _{\phi}+p_{\phi}>>\rho _{rad}+p_{rad}$, from Eq.~(\ref{27}) we obtain for $b$ the expression
\be
b+2\approx b\approx \sqrt{\frac{3}{4}}\frac{\dot{\phi}}{\sqrt{\rho _{rad}}}\tan 2\alpha .
\ee
We would like to point out that the magnitude of $b$ is determined by the ratio $\dot{\phi}/\sqrt{\rho _{rad}}=\sqrt{2\left[\rho _{\phi}-U(\phi)\right]/\rho_{rad}}$, and the condition can also be satisfied for small values of $\alpha$, especially by taking into account the fact that at the beginning of the warm inflationary era the energy density of the radiation is small.  Thus within the framework of these approximations we obtain first for $F$ the expression
\be
F\approx \frac{2}{3}b^2\rho _{rad}=\frac{1}{2}\dot{\phi}^2\tan ^2 2\alpha.
\ee
Now, by assuming that the angle $\alpha$ is small, we will approximate $\tan 2\alpha \approx 2\alpha$, thus obtaining
\be
F\approx 2\alpha ^2\dot{\phi}^2,
\ee
where generally $\alpha $ is a function of the thermodynamic parameters of the scalar field and of the radiation fluid, $\alpha =\alpha \left(\rho _{\phi}, p_{\phi},\rho _{rad}, p_{rad}\right)$. Then, under this assumptions and simplifications the energy balance equations (\ref{61}) and (\ref{62}) take the form
\be\label{70}
\ddot{\phi}+3H\left[1+\frac{8\alpha \dot{\alpha}}{3H\left(1+8\alpha ^2\right)}\right]\dot{\phi}+\frac{U'(\phi)}{1+8\alpha ^2}=0,
\ee
\be\label{71}
\dot{\rho} _{rad}+4H\rho _{rad}=4\alpha ^2\dot{\phi}^2\left(\frac{\dot{\alpha}}{\alpha}+\frac{\ddot{\phi}}{\dot{\phi}}\right)+12\alpha ^2\dot{\phi}^2H+2\alpha^2\dot{\phi}^2H_3.
\ee

To obtain the full picture of the cosmological evolution we need to also solve the gravitational Einstein equations. For the adopted form of the cosmological anisotropic warm inflationary energy-momentum tensor from Eqs.~(\ref{9}) it follows that we can assume $a_1(t)=a_2(t)$ without any
loss of generality. The gravitational field equations (\ref{8})-(\ref{10}) describing the noncomoving evolution of the scalar field - radiation fluid mixture take the form
\be\label{73a}
3\dot{H}+\sum _{k=1}^3H_k^2=-\left[\left(1+2\alpha ^2\right)\dot{\phi}^2+\rho _{rad}-U(\phi)\right],
\ee
\be\label{74a}
\frac{1}{V}\frac{d}{dt}\left(VH_i\right)=\alpha ^2\dot{\phi}^2+U(\phi)+\frac{1}{3}\rho_{rad},i=1,2,
\ee
\be\label{75a}
\frac{1}{V}\frac{d}{dt}\left(VH_3\right)=U(\phi)+\frac{1}{3}\rho _{rad}.
\ee

The expansion parameter $\theta =3H$ of the scalar field-radiation fluid filled Universe is
obtained as
\begin{equation}  \label{th}
\theta =2\frac{\dot{a}_1}{a_1}+\frac{\dot{a}_3}{a_3}.
\end{equation}
For the considered  warm inflationary anisotropic Universe the shear scalar $\sigma $  is given by
\begin{equation}  \label{sh}
\sigma =\frac{1}{\sqrt{3}}\left(\frac{\dot{a}_3}{a_3}-\frac{\dot{a}_1}{a_1}%
\right).
\end{equation}

With the use of Eqs.~(\ref{th}) and (\ref{sh}) it is possible to determine the
directional Hubble parameters $H_1$ and $H_3$ as functions  of the observationally detectable
cosmological parameters in the form
\begin{equation}
H_1=\frac{\theta }{3}-\frac{1}{\sqrt{3}}\sigma,\qquad H_3=\frac{\theta }{3}+%
\frac{2}{\sqrt{3}}\sigma,
\end{equation}
respectively.  The difference of the anisotropic cosmological
pressures is obtained as
\begin{equation}
\Psi -\Pi =\sqrt{3}\sigma +\theta \sigma.
\ee

The volume evolution Eq.~(\ref{41}) can be written as
\be
\ddot{V}-3\left(U(\phi)+\rho_{rad}+\frac{2}{3}\alpha ^2\dot{\phi}^2\right)V=0.
\ee

\section{The slow roll approximation}\label{sect4}

In the present Section we will investigate the warm inflationary scenario introduced in the previous Section  by assuming both the quasi stability and the slow roll conditions. Besides these two well known assumptions, it is usual to study the warm inflationary models in two important limits, namely the weak and the strong dissipative regimes, respectively. Here at first we obtain the general forms of the relevant equations describing the warm inflationary regime. Then we will discuss in detail the asymptotic regimes of the inflationary behavior in the presence of noncomoving motions of scalar field and radiation

 \subsection{Evolution equations for noncomoving warm inflation}

 We begin our investigation with the wave equation, i.e. Eq.~(\ref{70}), in which, by assuming the condition $\ddot{\phi}/\dot{\phi}\ll3H$, the time variation $\dot{\phi}$  of the inflaton field can be expressed as
\be\label{dotphi}
\dot{\phi}=-\frac{U'(\phi)}{ 3H(1+Q)(1+8\alpha ^2)},
\ee
where
 \begin{equation}\label{Dissipation-Q}
 Q=\frac{\Gamma}{3H}=\frac{8\alpha \dot{\alpha}}{3H\left(1+8\alpha ^2\right)}\,,
   \end{equation}
   gives the dissipation function, while, as already indicated in Eq.~(\ref{KG0}), $\Gamma$ is the dissipation coefficient. Now whereas we are working in a quasi stable environment, with  $\dot\rho_{rad} \ll H\rho_{rad}$ , Eq.~(\ref{71}) can be expressed as follows
\be\label{Density}
4H\rho _{rad}=4\alpha ^2\dot{\phi}^2\left(\frac{\dot{\alpha}}{\alpha}+3H+\frac{H_3}{2}\right).
\ee
From Eqs.~(\ref{74a}) and (\ref{75a}) one obtains immediately for the term $\alpha ^2\dot{\phi}^2$ the expression
\be\label{alphaphidot}
\alpha ^2\dot{\phi}^2=\frac{1}{V}\frac{d}{dt}\left(VH_1\right)-\frac{1}{V}\frac{d}{dt}\left(VH_3\right).
\ee

To find an explicit expression of the above equations, and without losing any generality,  one can assume that $a_3(t)=a_i^\lambda(t)$, $i=1,2$, in which $\lambda$ is a constant whose value should be obtained from the in-depth comparison with the data provided by the cosmological observations \cite{Ba1,Ba2,Naderi:2018kre}.
 Since the scale factor for the $x$  direction is the same as for the $y$ one, so that $a_1(t)=a_2(t)$, consequently, from Eqs.~ (\ref{55a}) and (\ref{56a}) we obtain at once
\begin{equation}\label{55aa}
V=a_{i}^{2+\lambda},\qquad H_{3}=\lambda \frac{\dot{a}_{i}}{a_{i}},\qquad i=1,2,
\end{equation}%
and
\begin{equation}\label{56aa}
H=\frac{1}{3}\left( 2+\lambda\right) H_{i}=\frac{\dot{V}}{3V}, \qquad i=1,2,
\end{equation}%
respectively. By combining Eqs.~(\ref{alphaphidot}) -(\ref{56aa}), it follows that the term $\alpha ^2\dot{\phi}^2$ can be expressed as follows,
\be\label{alphaphidot-Hi}
\alpha ^2\dot{\phi}^2=(1-\lambda)\big[(2+\lambda)H_i^2+\dot{H_i}\big], \qquad i=1,2.
\ee

Next in Eq.~(\ref{Density}) we suppose that
\be\label{alpha-Hi}
\frac{\dot{\alpha}}{\alpha}=-h_0H_i,  \qquad i=1,2,
 \ee
 where $h_0$ is a constant that can be constrained by the data.

 To determine $\alpha$ we can use the generalized de Sitter scale factor, the so called intermediate scale factor \cite{Ba1,Ba2,Naderi:2018kre},  as well as a power law expression of the scale factor. As the first example we can consider ${a}_i(t)=a_{0i}e^{\gamma t^n}$, while for the power law case we will introduce the scale factor as $\bar{a}_i(t)=\bar{a}_{0i}t^m$ (see \cite{Ghadiri:2018nok}, and references therein). Here $a_{0i},~\bar{a}_{0i},~n$ and $m$ are some constants that will be fixed by means of observations. From Eq.~(\ref{alpha-Hi}), and for the generalized de Sitter and power law cases, one finds,
\be\label{alphadeSitter-Hi}
\alpha=\alpha_0e^{-h_0\gamma t^n}=\alpha_0\left(\frac{a_i(t)}{a_{0i}}\right)^{-h_0}, \qquad i=1,2,
\ee
and
\be\label{alphapowerlaw-Hi}
\alpha=\bar{\alpha}_0t^{-h_0m}=\bar{\alpha}_0\left(\frac{\bar{a}_i(t)}{\bar{a}_{0i}}\right)^{-h_0}, \qquad i=1,2,
\ee
respectively, where ${\alpha}_0$ and $\bar{\alpha}_0$ are some arbitrary integration constants. From the above equations and from Eq.~(\ref{alphaphidot-Hi}) the expression of the inflaton field as a function of the cosmic time can be straightforwardly obtained. Thus,  for the generalized de Sitter and power law scale factors, respectively,  it follows that
\be\label{phi-deSitter}
\dot{\phi}^2=\frac{n\gamma(1-\lambda)}{\alpha_0^2e^{-2h_0\gamma t^n}}\big[(2+\lambda)t^n- n\gamma(1-n)\big]t^{n-2},
\ee
and
\be\label{phi-powerlaw}
\dot{\phi}^2=\frac{m(1-\lambda)}{\bar{\alpha}_0^2}\big[(2+\lambda)m- 1\big]t^{-2(1-h_0m)},
\ee
respectively.

In the warm inflationary scenarios, the slow-roll approximations are still legitimate, that is, {\it the rate of variation of the Hubble parameter during a Hubble time interval is assumed to be  smaller than unity}. This condition can be implemented in the cosmological model by using the behavior of the first slow-roll parameter
\begin{equation}\label{epsilon}
\epsilon_1 = - {\dot{H} \over H^2},
\end{equation}\
which satisfies the constraint $\epsilon _1<<1$. Besides the quasi de Sitter assumption leading to Eq.~(\ref{epsilon}), the energy density of the inflaton field is much larger as compared to the radiation energy density.  Moreover, the kinetic energy component of the total energy of the scalar field is insignificant as compared to its potential energy, {\it i.e.}  $\rho_\phi \gg  \rho_{rad}$ and $\rho_\phi \simeq U(\phi)$.

Then, from Eqs.~\eqref{w04},  \eqref{10}, \eqref{Density}, \eqref{56aa}, and \eqref{alpha-Hi}, it follows that
\begin{eqnarray}
  &3H^2 =\frac{(2+\lambda)^2}{3}H_i^2 \simeq  U(\phi)\;, \qquad i=1,2, \label{vfriedmann} \\
 &  \rho_{rad} :=  C_\gamma T^4 = \left(\frac{3}{2+\lambda}\right) \left(\frac{2h_0+3\lambda+4}{2}\right)\alpha^2\dot\phi^2\;, \label{radiationtemp}
\end{eqnarray}
where  by $T$ we have denoted the temperature of the radiation fluid. By substituting  Eq.~\eqref{epsilon} into \eqref{vfriedmann}, and with the use of Eq.~\eqref{dotphi}, it follows that in the warm inflationary model with noncomoving scalar field and radiation the slow-roll parameters are given by
\begin{equation}\label{vsrp}
\epsilon_1 = {1 \over 2(1+Q)} {U^{\prime 2}(\phi) \over {\left(1+8\alpha^2\right)U^2(\phi)}}\;, \quad \epsilon_2 = {\dot{\epsilon}_1 \over H \epsilon_1}\;.
\end{equation}

In an isotropic background, {\it i.e.} $\lambda=1$, and when $\alpha$ tends to zero, the first slow roll parameter goes back to the usual warm inflation relation \cite{w57,Sheikhahmadi:2019gzs,Rasheed:2020syk}.
To examine the effectiveness of a theoretical model, one basic approach is to contrast its predictions with the observations. In doing so for the noncomoving warm inflationary model, at first we will obtain some important perturbations parameters, including the amplitude of the scalar perturbations, $\mathcal{P}_s$, the amplitude of the tensor perturbations $\mathcal{P}_t $, the scalar and tensor spectral indices  $n_s,~n_t$, and the tensor-to-scalar ratio $r$, respectively. Then we will compare the predictions of the theoretical model with the recently released Planck-2018 data. According to the results of  \cite{w4,w7a,w22,w38}, the amplitude of the scalar perturbations is obtained in the form
\begin{equation}\label{pswarm}
\mathcal{P}_s = \left( { H^2 \over 2\pi \dot\phi } \right)^2 \left(1 + 2n_{BE} + \frac{2\sqrt{3}\pi Q}{\sqrt{3+4\pi Q}}{T \over H}\right) G(Q)\;,
\end{equation}
where by $n_{BE}$ we have denoted the Bose-Einstein distribution function, which is given by the expression $n_{BE}= \left[ \exp(H/T_{\delta\phi}) - 1 \right]^{-1}$, and where $T_{\delta\phi}$ is the temperature of the inflaton field fluctuations \cite{w38}. The function $G(Q)$, which describes the growth of the cosmological fluctuations, depends on $Q$ only, and it has arisen from the coupling of the inflaton scalar field and the radiation fluid \cite{w22,w38}.

Moreover, the scalar spectral index and its running, $\alpha_s$, can be determined from the amplitude of the scalar perturbations, and they are specified according to the definitions
\begin{equation}\label{nsaswarm}
n_s - 1 = {d\ln(\mathcal{P}_s) \over d\ln(k)}\;, \qquad \alpha_s = {d n_s \over d\ln(k)}\;.
\end{equation}

Tensor perturbations, representing the generation of the gravitational waves, are determined with the help of the  tensor-to-scalar ratio parameter $r=\mathcal{P}_t / \mathcal{P}_s$. For the amplitude of the tensor perturbations we obtain \cite{w22}
\begin{equation}\label{ptwarm}
\mathcal{P}_t = {2 H^2 \over \pi^2}\;.
\end{equation}

Another important observational quantity,  the tensor spectral index is constructed as
\begin{equation}\label{ntwarm}
n_t = {d\ln(\mathcal{P}_t) \over d\ln(k)}\;.
\end{equation}

To measure the quantity of cosmic expansion during  the inflationary dynamical evolution  it is customary to use the number of e-folds $\mathcal{N}$, which are defined according to
\begin{equation}\label{efoldf}
\mathcal{N} = \int_{t_\star}^{t_e} H \; dt = \int_{\phi_\star}^{\phi_e} {H \over \dot\phi} \; d\phi~,
\end{equation}
where the subscript $\star$ denotes the horizon exit values.

In the warm inflationary approach, it is a common practice to investigate the model in two different and important thermal approximations, which are known  as the strong and the weak dissipative regimes, respectively. In these two regimes the dissipative ratio satisfies the conditions  $Q \gg 1$, and $Q \ll 1$, respectively.  From Eq. \eqref{vsrp} it should be noted  that inflation ends when
\begin{equation}\label{endofinflation}
\epsilon_1 = 1+Q\;,
\end{equation}
where for the weak and the strong regimes $\epsilon _1$ behaves as $\epsilon_1 \simeq 1$, and $\epsilon_1 \simeq Q$, respectively \cite{Dimopoulos:2019gpz}.
In the subsequent Subsections, we will investigate in detail the noncomoving warm inflationary  model in the presence of scalar field and radiation fluid in these two limiting regimes.

\subsection{Noncomoving warm inflation in the weak dissipative regime}\label{Secweaka}

As we have already mentioned, the main characteristic of the weak dissipative regime is that the dissipative ratio is much smaller than unity, that is, $Q$ satisfies the condition $Q\ll1$ $(\Gamma \ll 3H)$. Consequently, we also have $(1+Q) \simeq 1$. Moreover, at the time of horizon crossing, in the weak dissipation approximation, the parameter $G(Q) \simeq 1$. Hence, from Eq.~\eqref{pswarm}, it follows that the amplitude of the scalar perturbations in the weak regime reads
\begin{equation}\label{psweak}
\mathcal{P}_s =2\mathcal{{P}}_{s0}  \; \left( { H^2 \over 2\pi \dot\phi } \right)^2 {T \over H}\;,
\end{equation}
where $\mathcal{{P}}_{s0}$ is a constant should be determined comparing to data.
To investigate the  consistency of the warm inflationary models one has to test the condition $T/H>1$,  besides other observational constraints. Thus, for the case of the noncomoving warm inflation from Eqs.~\eqref{dotphi}, \eqref{vfriedmann} and \eqref{radiationtemp} we obtain
\begin{equation}\label{THweak}
{T \over H}=\left(\frac{\sqrt{T_0}\alpha}{1+8\alpha^2}\right)^{1\over 2}\frac{U^{\prime\frac{1}{2}}}{U^{\frac{3}{4}}}\;,
\end{equation}
where $$T_0=\frac{3^2}{(2+\lambda)C_\gamma}\left(\frac{2h_0+3\lambda+4}{2}\right).$$ Hence the ratio of the photon temperature and of the Hubble function depends on the function $\alpha$, describing the differences in the four velocities of the radiation and scalar field.

From Eq.~\eqref{nsaswarm}, and from the combination of Eqs.~\eqref{psweak} and \eqref{THweak}, the scalar spectral index is obtained as
\be\label{nsweak}
 n_s -1= -{9 \over 2} \; \epsilon_1 + {3} \; \eta -  \beta\;,
\ee
where
\begin{equation}\label{epsilonweal}
\epsilon_1 = {1 \over 2(1+8\alpha^2)} {U^{\prime 2}(\phi) \over {U^2(\phi)}}\;.
\end{equation}

For the potential slow roll parameter $\eta$ we find
\begin{equation}\label{etaweak}
\eta= {1 \over 2(1+8\alpha^2)} {U^{\prime\prime}(\phi) \over {U(\phi)}}\;,
\end{equation}
while the slow-roll parameter $\beta$ can be obtained as
  \begin{equation}\label{betaweak}
\beta={1 \over 2(1+8\alpha^2)} {U^{\prime}(\phi)\Gamma'_{eff} \over {U(\phi)\Gamma_{eff}}}\;,
\end{equation}
where $\Gamma_{eff}=\alpha(1+8\alpha^2)^3$.

At the time of horizon crossing,  from Eqs.~\eqref{dotphi}, \eqref{ptwarm} and \eqref{psweak} it follows that the tensor-to-scalar ratio is given by
\begin{equation}\label{rweak}
r =  \frac{8\epsilon_1}{1+8\alpha^2} {H \over T}\;.
\end{equation}

As one can see, all these important observational parameters are functions of $\alpha$, describing the deviations from the comoving motion of the fluid components of the early Universe.

\subsubsection{Weak dissipative regime with power law scale factor}\label{Secweaka-powlaw}

In this Subsection we are going to investigate the compatibility of  the theoretical results of the noncomoving warm inflationary model obtained within the weak dissipative regime with power law scale factor to the recent observational data. In doing so we will consider Planck 2013, 2015, and 2018 data sets as our criteria.

From Eq.~\eqref{phi-powerlaw} a relation for the cosmic time as a function of the scalar field can be obtained as
\begin{equation}\label{tphiweak-plaw}
t(\phi)=\left[\frac{{h_0} \bar{\alpha}_0  m  }{\sqrt{{m(1-\lambda )  \big[(\lambda +2) m-1\big]}}}\phi \right]^{\frac{1}{{h_0} m}}\, .
\end{equation}

By substituting Eq.~\eqref{tphiweak-plaw} into Eqs.~\eqref{dotphi} and \eqref{phi-powerlaw},  for the weak dissipative regime with power law scale factor,  the scalar field potential is given as a function of the scalar field by
\begin{eqnarray}\label{Potentialweak-plaw}
\hspace{-0.5cm}&&U(\phi)=  \Bigg[ 3h_0^2 m \left(1-mh_0\right)\phi ^2
+24 (\lambda -1) (h_0 m-1) \times \nonumber\\
\hspace{-0.5cm}&&(m \lambda +2m-1)\Bigg]
\times\frac{h_0 m^3 \left[\frac{{h_0} m \phi }{\sqrt{-\frac{\left(\lambda -1\right) m \left((\lambda +2) m-1\right)}{\bar{\alpha}_0 ^2}}}\right]^{-\frac{3}{{h_0} m}}\phi}{{\left(h_0 m-3\right) \left(h_0 m-1\right)}} .\nonumber\\
\end{eqnarray}

In addition, by taking Eqs.~\eqref{epsilonweal} equal to unity, and after some simple mathematical manipulations, we obtain the magnitude of the scalar field at the end of noncomoving warm inflation as presented in Table~\ref{Tab1a}.
\begin{widetext}
\begin{table*}[t]
  \centering
\begin{tabular}{|c|c|c|c|c|c|c|c|c|c|c|}
\hline
\hline
$m$ & $\lambda$ & $\phi_\star$ & $\phi_{\rm e}$ & $h_0$ & $\bar{\alpha}_0$ & $n_{\rm s}$ & $r_\star$ & $\mathcal{P}_{s0}$ & $T(\phi_\star)/H(\phi_\star)$ & $\mathcal{N}$\\
\hline
$29.950$ & $11.810$  & $1809.87$ & $3169.22$ & $0.0096592$& $26.1$ & $0.944287$ &$0.0000111051$ & $2.96471$ & $1.05881$ &$58$\\
\hline
   $29.952$ & $11.812$ & $1792.78$ & $3169.75$ & $0.0096591$ & $26.02$ & $0.938879$ &$0.0000111068$ & $2.52561$ & $1.01903$ &$59$\\
   \hline
   $29.949$ & $11.814$ & $1793.13$ & $3170.36$ & $0.00965899$ & $26.04$ & $0.964083$ &$0.0000110715$ & $2.56107$ & $1.02204$&$59$ \\
   \hline
   $29.948$ & $11.816$ & $1810.85$ & $3170.9$ & $0.00965898$ & $26.06$ & $0.968554$ &$0.0000111356$ & $2.92386$ & $1.05495$&$58$ \\
   \hline
   \hline
  \end{tabular}
\caption{Estimation of the free parameters of the noncomoving warm inflationary model in the weak dissipation regime with power law scale factors by using  the Planck 2013, 2015 and 2018 data sets, and for $C_\gamma=50$. To obtain the presented results we have fixed the power spectrum to its horizon exit value, i.e., $\mathcal{P}_s=2.17\times 10^{-9}$. For this case we restrict our analysis to real values of the scalar field at the horizon exit.}\label{Tab1a}
\end{table*}
\end{widetext}

\begin{figure*}[tb]
\centering
\includegraphics[scale=0.81]{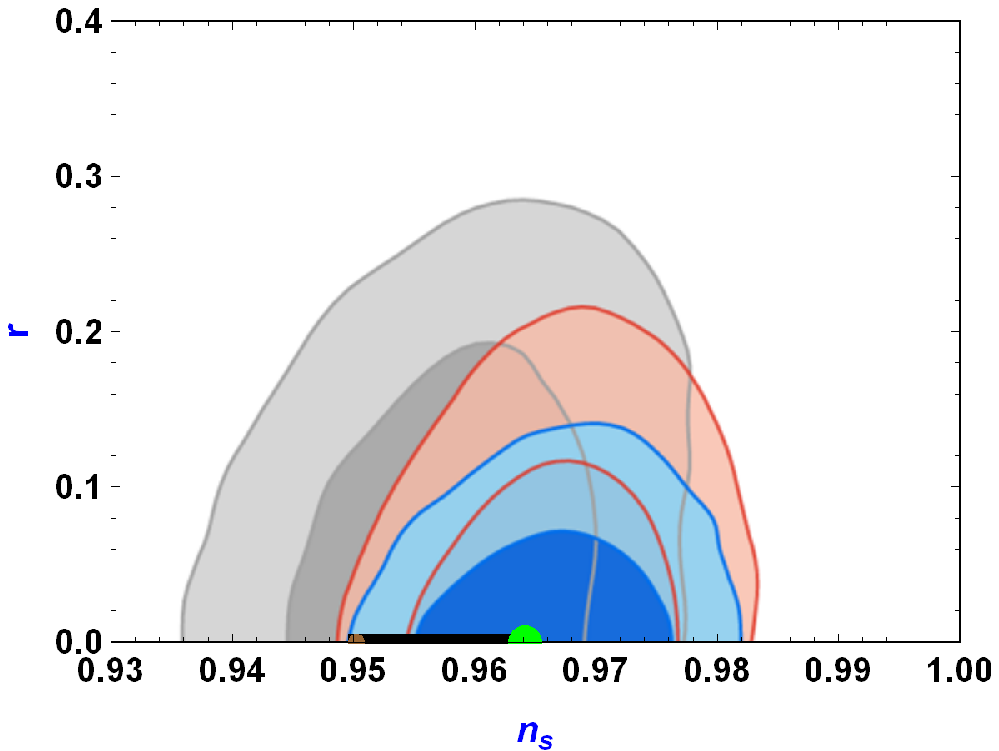}
\includegraphics[scale=0.81]{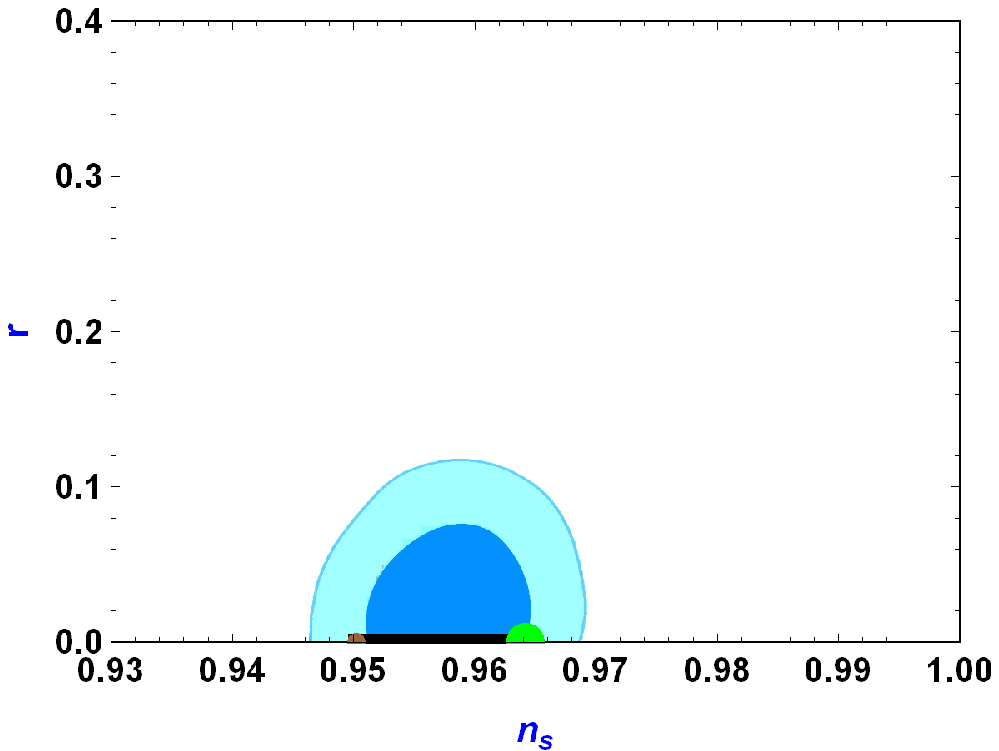}
\caption{
 The $r-n_s$ diagram comparing the predictions of the noncomoving warm inflation model in the weak regime with power law scale factors, having the model parameters of Table \ref{Tab1a}, with the observational data of the Planck $2013$, $2015$ and 2018 data sets, respectively. In the left panel, the likelihoods  of  the Planck 2013 data are depicted with grey contours, of the Planck TT+lowP data with red contours, and of the Planck TT,TE,EE+lowP (2015) data with blue contours. In the right panel, the results from the Planck 2018 data are plotted by dark and light blue colors, indicating the $68\%$ and $95\%$ confidence levels, respectively. In both Figures the thick black lines represent the predictions of the noncomoving warm inflation model, while the small, brown, and large, green, circles are the values of $n_{\rm s}$ at the number of e-folds $\mathcal{N}=58$ and $\mathcal{N}=62$, respectively.}
\label{r-nsweakpowerlaw}
\end{figure*}

\begin{figure}[tb]
\includegraphics[scale=.67]{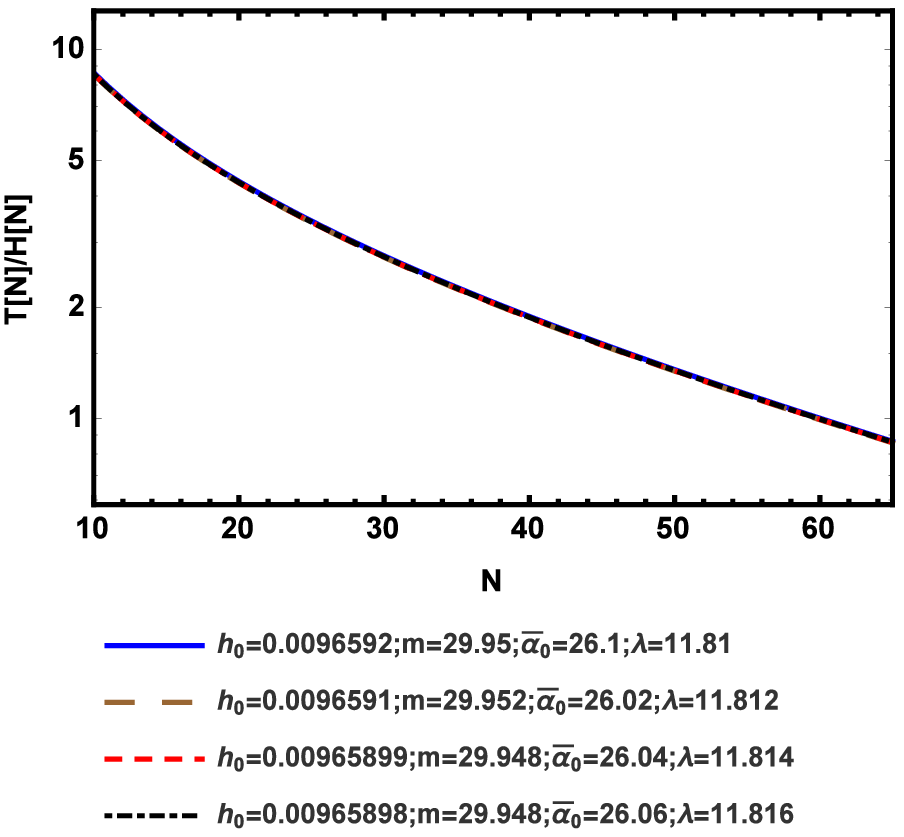}
\caption{
 The ratio of the temperature and of the Hubble parameter in the course of the  evolution of the noncomoving warm inflationary model in the weak dissipation regime, with power law scale factors,  as a function  the number of e-folds, for different numerical values of the parameters of the model. During the inflationary era the temperature of the Universe is bigger than the Hubble parameter, and their ratio significantly decreases when approaching the end of warm noncomoving inflation.}
\label{THweakpowerlaw}
\end{figure}

Then, by using Eqs.~\eqref{phi-powerlaw} and  \eqref{efoldf}, the scalar field at the horizon exit can be obtained in the noncomoving warm inflation model as
\begin{equation}\label{phistarweak-plaw01}
\phi_{\star}=\phi_e\times \exp[-h_0\mathcal{N}].
\end{equation}

Now, by using Eq.~\eqref{phistarweak-plaw01}, and the results of the Subsection \ref{Secweaka}, especially Eqs.~\eqref{nsweak} and \eqref{rweak}, we can examine the behavior of the scalar spectral index versus the tensor to scalar ratio, and of the temperate-Hubble parameter ratio against the number of e folds at the horizon exit in the noncomoving warm inflationary model.

The results of the comparison between the theoretical results and the observations can be summarized in Table~\ref{Tab1a}, and Figs.~\ref{r-nsweakpowerlaw} and \ref{THweakpowerlaw}, respectively.


\subsubsection{Weak dissipative regime with generalized de Sitter  scale factor}\label{Secweaka-powlaw}

Now we are going to investigate the compatibility of  the theoretical predictions of the noncomoving warm inflationary model, obtained within the weak dissipative regime, with  generalized de Sitter scale factor, to the recent observational data. In doing so we will consider the Planck 2013, 2015, and recent 2018 data sets as our criteria.

From Eq.~\eqref{phi-deSitter}, the relation of the cosmic time as a function of the scalar field can be obtained as
\begin{equation}\label{tphiweak-Interm}
t(\phi)=-\frac{1}{2 \gamma  {h}_0}\ln \left[-\frac{\lambda ^2+\lambda -2}{ \gamma \left({\alpha }_0 {h}_0 \phi\right) ^2}\right] .
\end{equation}

By substituting  Eq.~\eqref{tphiweak-Interm} into Eqs.~\eqref{dotphi} and \eqref{phi-deSitter},  for the weak regime with generalized de Sitter case,  the scalar field potential as a function of the warm inflationary scalar field reads
\begin{equation}\label{Potentialweak-Interm}
U(\phi)=\frac{3 \gamma  \left[8 \left(-\lambda ^2-\lambda +2\right) \ln \left(-\frac{\lambda ^2+\lambda -2}{{\alpha }_0^2 \gamma  {h}_0^2 \phi ^2}\right)+\gamma {h}_0^2 \phi ^2\right]}{2 {h_0}}\, .
\end{equation}

In addition, by taking Eqs.~\eqref{epsilonweal} equal to unity, and after some simple mathematical manipulations, we obtain the magnitude of the scalar field at the end of inflation as presented in Table~\ref{Tab1b}.
\begin{table*}[t]
  \centering
\begin{tabular}{|c|c|c|c|c|c|c|c|c|c|c|}
\hline
\hline
$\gamma$ & $\lambda$ & $|\phi_\star|$ & $|\phi_{\rm e}|$ & $h_0$ & ${\alpha}_0$ & $n_{\rm s}$ & $r_\star\times 10^{-7}$  & $\mathcal{P}_{s0}\times 10^{-20}$&  $T(\phi_\star)/H(\phi_\star)$ & $\mathcal{N}$\\
\hline
$33$ & $0.02$ & $0.00369826$ & $1.58425$ & $0.101$ & $60$ & $0.959775$ &$4.96773$ &  $1.26815$ & $3.3729$& $60$\\
\hline
$34$ & $0.03$ & $0.00351334$ & $1.5981$ & $0.102$ & $61$ & $0.958454$ &$4.76646$ &  $1.19188$ & $3.36302$& $60$\\
   \hline
$35$ & $0.04$ & $0.00334193$ & $1.61412$ & $0.103$ & $62$ & $0.956301$ &$4.73632$ &  $1.13156$ & $3.35017$& $60$\\
   \hline
 $36$ & $0.05$ & $0.00317758$ & $1.62965$ & $0.104$ & $63$ & $0.954101$ &$4.69175$ &  $1.07226$ & $3.33757$& $60$\\
\hline
\hline
  \end{tabular}
\caption{Estimation of the free parameters of the noncomoving warm inflationary model in the weak dissipation regime with generalized de Sitter  scale factors by using  the Planck 2013, 2015 and 2018 data sets, and for $C_\gamma=55$. To obtain the presented values of the model parameters we have fixed the power spectrum to its horizon exit value, i.e., $\mathcal{P}_s=2.17\times 10^{-9}$. For this case we restrict our analysis to real values of the scalar field at the horizon exit.}\label{Tab1b}
\end{table*}

\begin{figure*}[tb]
\centering
\includegraphics[scale=.81]{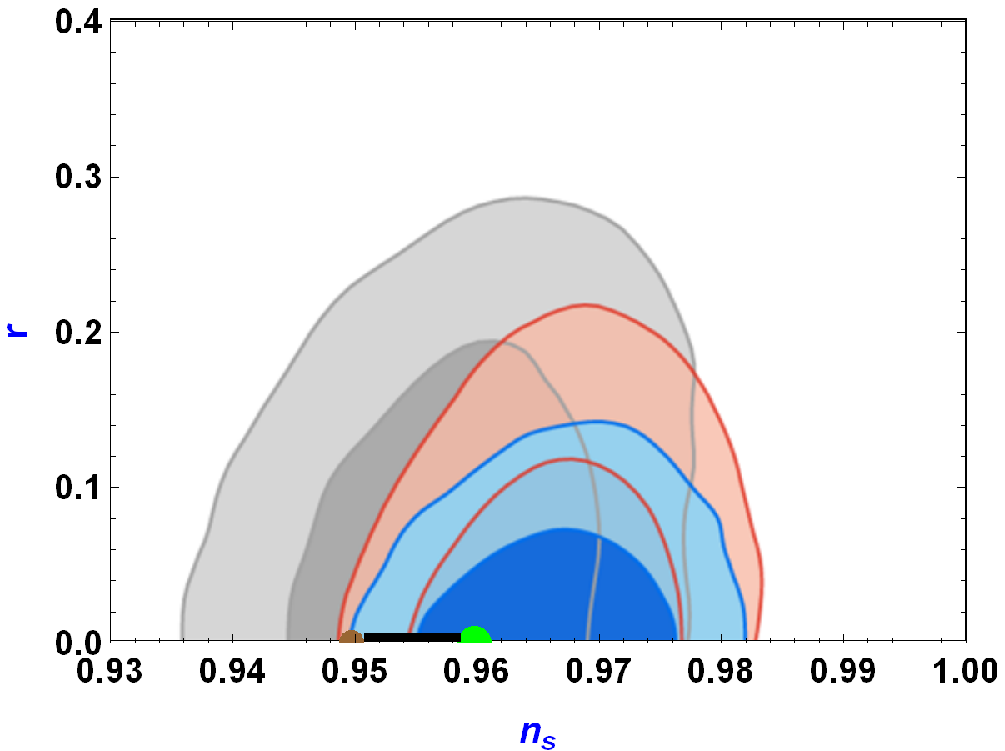}
\includegraphics[scale=.81]{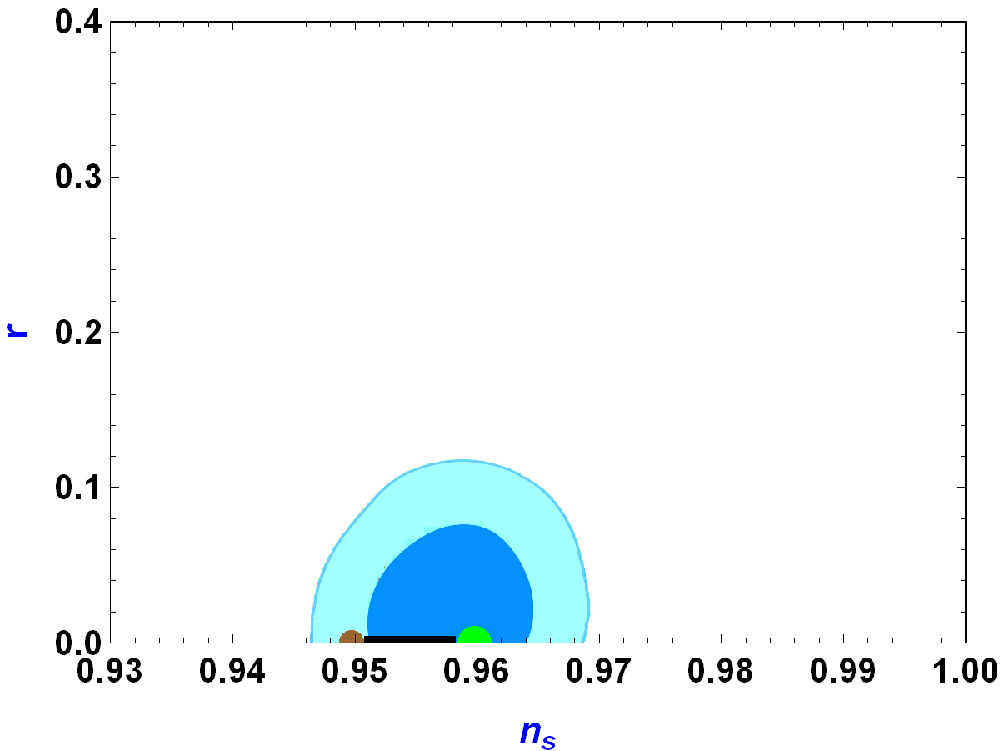}
\caption{
 The $r-n_s$ diagram comparing the predictions of the noncomoving warm inflation model in the weak regime with generalized de Sitter  scale factors, having the model parameters of Table \ref{Tab1a}, with the observational data of the Planck $2013$, $2015$ and 2018 data sets. In the left panel, the likelihoods  of  the Planck 2013 data sets are represented by grey contours, of the Planck TT+lowP data by red contours, while the Planck TT,TE,EE+lowP (2015) data are indicated by blue contours. In the right panel, the results of the Planck 2018 data are plotted in dark and light blue colors, referring to the $68\%$ and $95\%$ confidence levels, respectively. In both Figures, the thick black lines show the predictions of the noncomoving warm inflation model, while the small, brown, and large, green, circles are the values of $n_{\rm s}$ at the number of e-folds $\mathcal{N}=58$ and $\mathcal{N}=62$, respectively.}
\label{r-nsweakIntermddiate}
\end{figure*}
\begin{figure}[tb]
\includegraphics[scale=.67]{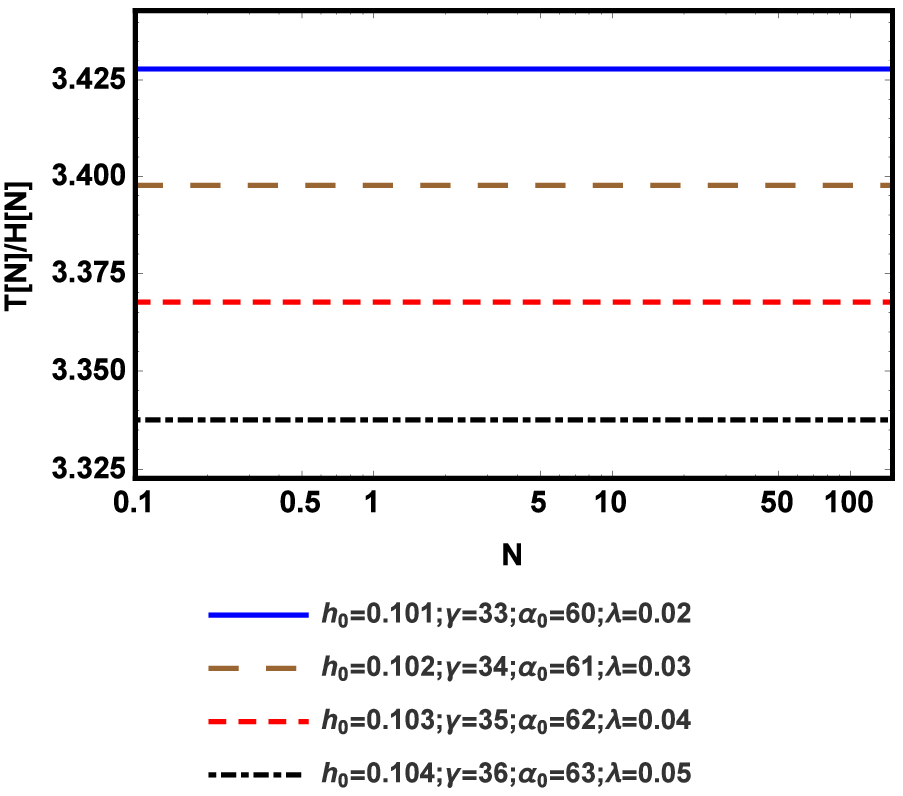}
\caption{
 The ratio of the temperature and of the Hubble parameter in the course of the nocomoving warm inflationary evolution in the weak dissipative regime with generalized de Sitter  scale factors as a function of the number of e-folds, for different numerical values of the parameters of the model.}
\label{THweakintermediate}
\end{figure}

Then, by using Eqs.~\eqref{phi-deSitter} and  \eqref{efoldf}, the scalar field at the horizon exit can be obtained in the noncomoving warm inflation model as
\begin{equation}\label{phistarweak-IntMideiate01}
\phi_{\star}= \phi_e\times \exp[-h_0\mathcal{N}].
\end{equation}

Now, by using  Eq.~\eqref{phistarweak-IntMideiate01}, and by taking into account the results of the Subsection \ref{Secweaka}, especially Eqs.~\eqref{nsweak} and \eqref{rweak}, we can investigate the behavior of the scalar spectral index versus the tensor to scalar ratio, and of the temperate-Hubble parameter ratio against the number of e folds at the horizon exit for the noncomoving warm inflationary mode with generalized de Sitter scale factor.

The results of the comparison between the theoretical results and the observations have been summarized in Table~\ref{Tab1b}, and Figs.~\ref{r-nsweakIntermddiate} and \ref{THweakintermediate}, respectively.


  \subsection{Strong dissipative regime}\label{Secstronga}

  In the strong dissipative regime, Eq.~\eqref{pswarm} takes the form
\begin{equation}\label{psstrong-GQ}
  \mathcal{P}_s = \left( { H^2 \over 2\pi \dot\phi } \right)^2 \; \sqrt{3\pi Q} \; {T \over H} \times G(Q) \;.
\end{equation}

In the following we will consider models in which the dissipation coefficient is introduced phenomenologically via an ansatz of the form $\Gamma= \Gamma_0 T^{\zeta}/ \phi^{\zeta-1}$, where $\Gamma _0$ and $\zeta $ are constants \cite{w22,w38,w56,Rasheed:2020syk}. Then the function $G(Q)$ that depends on the different values of the parameter $\zeta$, can be approximated as \cite{w22,w38}
\begin{eqnarray*}
\zeta=1  & \longrightarrow & G(Q)\simeq 1+ 0.127 Q^{4.330}+ 4.981Q^{1.946},\\
\zeta=3  & \longrightarrow & G(Q)\simeq 1+ 0.0185Q^{2.315}+ 0.335 Q^{1.364},\\
\zeta=-1 & \longrightarrow & G(Q)\simeq \frac{1+ 0.4 Q^{0.77}}{(1+0.15Q^{1.09})^2}.
\end{eqnarray*}

 Hence all the necessary parameters required to study the evolution of a warm inflationary evolution come from the above expressions of the decay rate $\Gamma$.  However, there is no need to restrict our study to only these three parameters. But, without any loss of generality, and in order to check the compatibility of the noncomoving warm inflationary model with the observations, following \cite{w56}, we only consider one of the $\zeta $ values per each case.

Therefore, in a more convenient way, and by taking into account that $Q\gg 1$,  we can write down the expressions of the function $G(Q)=a_\zeta Q^{b_\zeta}$, where
 \begin{eqnarray*}
\zeta=1   & \longrightarrow &  a_\zeta=0.127, \quad  b_\zeta={4.330},\\
\zeta=3   & \longrightarrow &  a_\zeta=0.0185, \quad  b_\zeta={2.315},\\
\zeta=-1  & \longrightarrow &  a_\zeta =17.78, \quad  b_\zeta={-1.41}\;.\\
\end{eqnarray*}

Therefore the amplitude of the scalar perturbations is obtained as
\begin{equation}\label{psstrong}
  \mathcal{P}_s = \mathcal{{P}}_{s0} \left( { H^2 \over 2\pi \dot\phi } \right)^2 \; \sqrt{3\pi Q} \; {T \over H} \times a_\zeta Q^{b_\zeta}\;.
\end{equation}

For the strong dissipation case, by taking into account Eqs.~\eqref{dotphi}, \eqref{vfriedmann}, and \eqref{radiationtemp}, we obtain
\begin{equation}\label{THstrong}
{T \over H}=\left(\frac{3\sqrt{T_0}\alpha}{1+8\alpha^2}\right)^{1\over 2}\frac{U^{\prime\frac{1}{2}}}{\Gamma^\frac{1}{2}U^{\frac{1}{2}}},
\end{equation}
where $T_0$ is the same as in Eq.~\eqref{THweak}. By taking the time derivative of  Eq.~\eqref{psstrong}, and using the definitions of Eq.~ \eqref{nsaswarm}, leads to the scalar spectral index in the strong dissipative regime of the noncomoving warm inflation as given by
\begin{eqnarray}\label{nsasstrong}
  n_s -1 &=&({b_\zeta}-\frac{5}{2} ) \epsilon_1 +{3} \eta-4 \beta-2b_\zeta \beta_1,
\end{eqnarray}
where
\begin{equation}\label{epsilonstrong}
\epsilon_1 = {\frac{1}{2Q(1+8\alpha^2)}}{U^{\prime 2}(\phi) \over {U^2(\phi)}}.
\end{equation}

For the potential slow roll parameter $\eta$ we find
\begin{equation}\label{etastrong}
\eta= {\frac{1}{2Q(1+8\alpha^2)}} {U^{\prime\prime}(\phi) \over {U(\phi)}}.
\end{equation}

The slow-roll parameters $\beta$ and $\beta_1$ are given as,
  \begin{equation}\label{betastrong}
\beta={\frac{1}{2Q(1+8\alpha^2)}}{U^{\prime}(\phi)\Gamma' \over {U(\phi)\Gamma}}\;,
\end{equation}
and
  \begin{equation}\label{beta1strong}
\beta_1={\frac{1}{2Q(1+8\alpha^2)}}{U^{\prime}(\phi)\Gamma'_{eff} \over {U(\phi)\Gamma_{eff}}}\;,
\end{equation}
respectively, where $\Gamma_{eff}=\alpha\Gamma$.  We notice at this moment that the function $G(Q)$ for $\zeta=1$ and $\zeta=3$  has an acceptable behavior for the value  $\zeta=-1$, or even for $\zeta=1$.

The amplitude of the tensor perturbations in the strong dissipative regimes is obtained as \cite{w22,w38}
\begin{equation}\label{ptstrong}
\mathcal{P}_t = {2 H^2 \over \pi^2}.
\end{equation}

Then, from Eqs.~\eqref{psstrong-GQ},  \eqref{epsilonstrong} and \eqref{ptstrong}, the tensor-to-scalar ratio reads
\begin{equation}\label{rstrong}
r =\frac{16\epsilon_1\sqrt{ Q}}{\sqrt{3\pi} \; (1+8\alpha^2) G(Q)} {H \over T}\;.
\end{equation}

\subsubsection{Strong dissipative regime with power law scale factor}\label{SecStronga-powlaw}

Following the procedure of the Subsection~\ref{Secweaka},  we are going to investigate the accuracy of the noncomoving warm inflation model with power law scale factor in the strong regime. First, by substituting  Eq.~\eqref{tphiweak-plaw} into Eqs.~\eqref{dotphi} and \eqref{phi-powerlaw},  for the strong regime with power law scale factor the  scalar field potential as a function of the scalar field is expressed as
\bea\label{Potentialstrong-Powelaw}
U(\phi)&=&4 h_0 (\lambda -1) m^2 \left[(\lambda +2) m-1\right] \times \nonumber\\
&&\left\{\frac{{h_0} m \phi }{\sqrt{-\frac{(\lambda -1) m \left[(\lambda +2) m-1\right]}{\bar{\alpha }_0^2}}}\right\}^{-\frac{2}{{h_0} m}}\, .
\eea

Also from Eqs.~\eqref{Potentialstrong-Powelaw} and \eqref{Dissipation-Q}, for the  dissipation ratio $Q$ one obtains
\begin{equation}\label{Dissipstrong-Powelaw}
Q(\phi)=-\frac{8 \left(\frac{{\bar{\alpha}_0} {h_0} m \phi }{\sqrt{(\lambda -1) (-m) ((\lambda +2) m-1)}}\right)^{\frac{1}{{h_0} m}}}{3 m \phi  \left(8-\frac{\text{h0}^2 m \phi ^2}{(\lambda -1) ((\lambda +2) m-1)}\right)}\, .
\end{equation}

Now we want to test  the theoretical predictions of the noncomoving warm inflationary model in the strong dissipative regime by comparing the theoretical results to the observational data sets. To perform such a comparison, at first and by means of Eqs.~\eqref{endofinflation} and \eqref{epsilonstrong}  for the end of inflation values of the scalar field, we obtain the results expressed in Table~\ref{Tab2a}.

\begin{table*}[t]
  \centering
  \centering
\begin{tabular}{|c|c|c|c|c|c|c|c|c|c|c|}
\hline
\hline
$m$ & $\lambda$ & $\phi_\star$ & $\phi_{\rm e}$ & $h_0$ & $\bar{\alpha}_0$ & $n_{\rm s}$ & $r_\star$ & $\mathcal{P}_{s0}\times 10^{-16}$ & $T(\phi_\star)/H(\phi_\star)$ & $\mathcal{N}$\\
\hline
$7.79931$ & $217$  & $0.102944$ & $13.8467$ & $0.100033$& $188$ & $0.954748$ &$0.0892411$ & $8.48726$ & $1.00639$ &$50$\\
\hline
   $7.79930$ & $218$ & $0.101816$ & $13.6944$ & $0.100032$ & $189$ & $0.954852$ &$0.087325$ & $8.05104$ & $1.00316$ &$50$\\
   \hline
   $7.79929$ & $216$ & $0.106635$ & $14.3446$ & $0.100035$ & $186$ & $0.954644$ &$0.0945849$ & $9.73916$ & $1.02192$&$50$ \\
   \hline
   $7.79928$ & $215$ & $0.107853$ & $14.5091$ & $0.100036$ & $185$ & $0.954538$ &$0.0967116$ & $10.02805$ & $1.02537$&$50$ \\
   \hline
   \hline
  \end{tabular}
\caption{The estimations of the free parameters of the noncomoving warm inflationary model in the strong dissipation regime with power law scale factor by using the Planck 2013, 2015 and 2018 data sets, and for $C_\gamma=50$. We have restricted our analysis to  real values of the scalar field at the horizon exit. To obtain the presented results we have fixed the power spectrum to its horizon exit value, i.e., $\mathcal{P}_s=2.17\times 10^{-9}$, and we have used  $a_\zeta=0.127$, and $ b_\zeta={4.330}$, respectively.}\label{Tab2a}
\end{table*}

\begin{figure*}[tb]
\centering
\includegraphics[scale=.81]{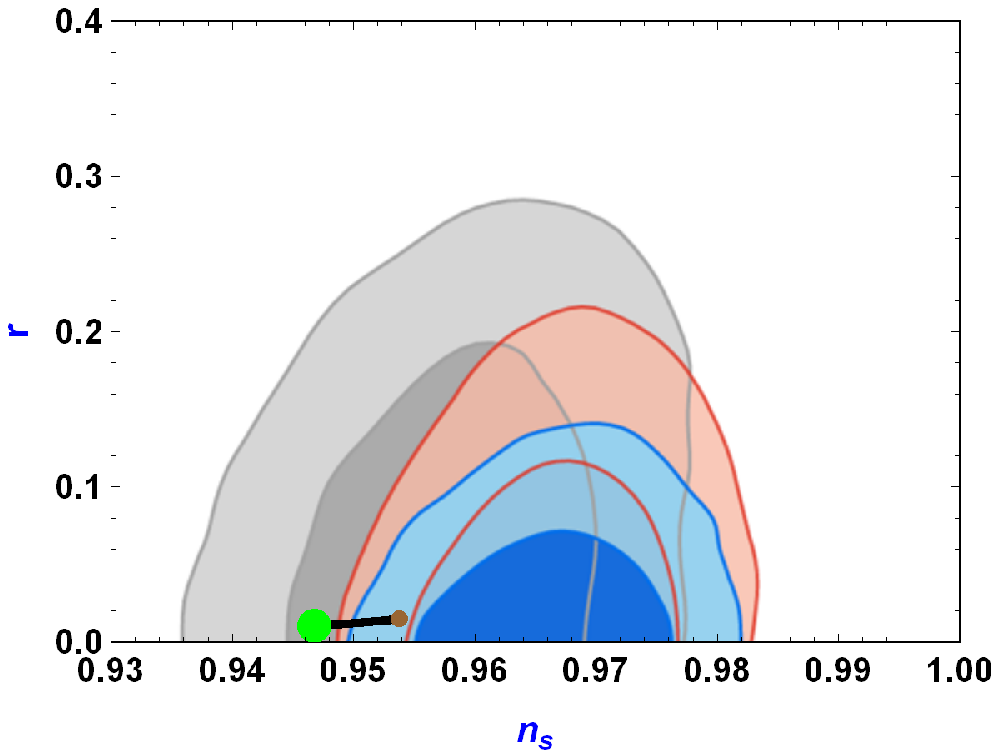}
\includegraphics[scale=.81]{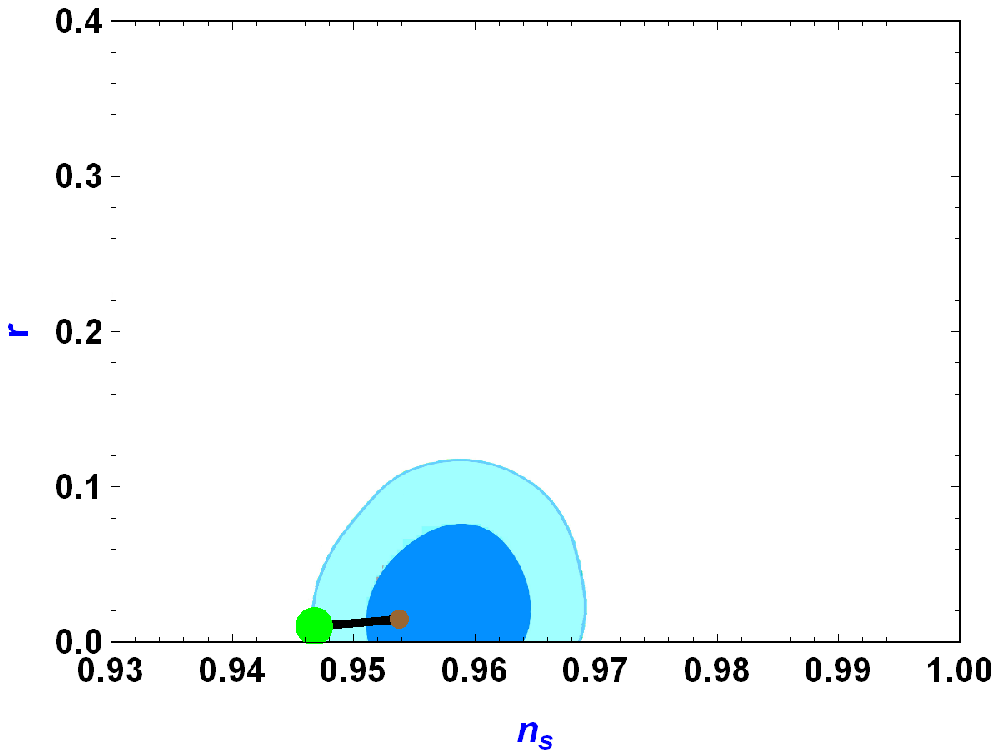}
\caption{
 The $r-n_s$ diagram comparing the predictions of the non-comoving warm inflation model in the strong dissipative regime  with power law scale factor and with the free parameters given in Table~\ref{Tab2a}, with the Planck $2013$, $2015$ and 2018 data sets. In the left panel, the likelihoods of  Planck 2013 data are represented with grey contours, of the Planck TT+lowP data are indicated with red contours, while the  Planck TT,TE,EE+lowP (2015) data are shown with blue contours. In the right panel, the recent results of the Planck 2018 data are depicted by dark and light blue colors, corresponding to $68\%$ and $95\%$ confidence levels, respectively. In both Figures the thick black lines represent the theoretical predictions of the noncomoving warm inflationary model, with the small, brown, and large, green, circles corresponding to the values of $n_{\rm s}$ at the number of e-folds $\mathcal{N}=50$ and $\mathcal{N}=55$, respectively.}
\label{r-nsStrongPowerlaw}
\end{figure*}
\begin{figure}[tb]
\includegraphics[scale=.61]{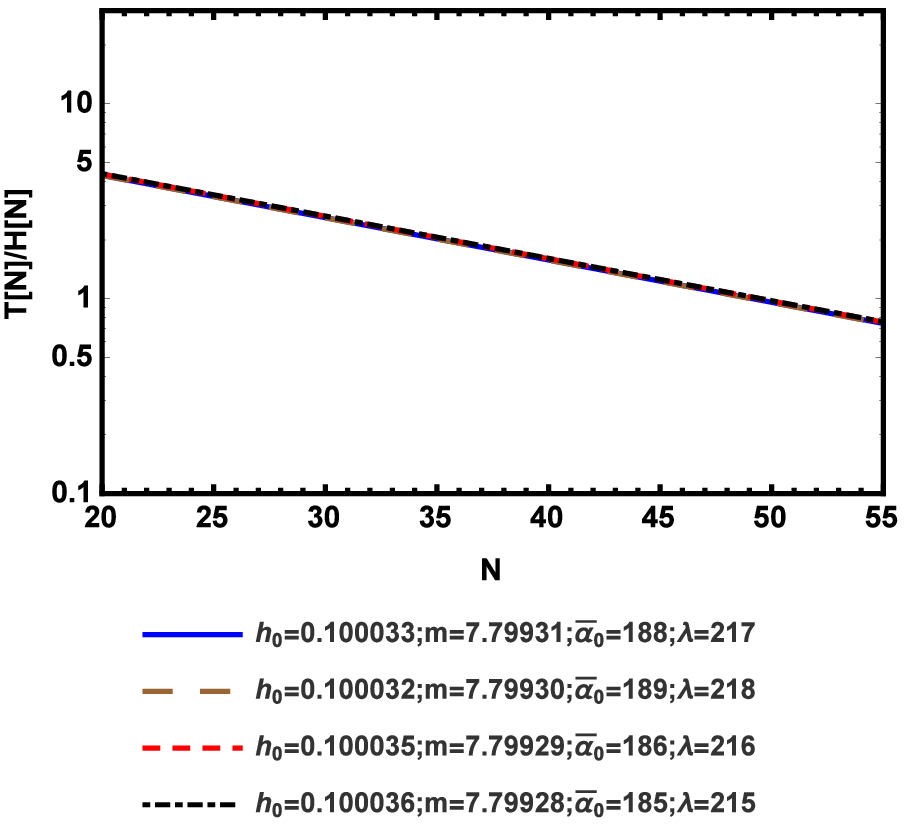}
\caption{
 The ratio of the temperature and of the Hubble parameter in the course of the inflationary evolution of the noncomoving warm inflationary model in the strong dissipative regime with power law scale factors as a function of the number of e-folds, and  for different numerical values of the parameters of the model.}
\label{THstrongpowerlaw}
\end{figure}

Consequently, by means of Eqs.~\eqref{phi-powerlaw} and  \eqref{efoldf}, the  scalar field at the horizon exit for the strong dissipative regime has the value
\begin{equation}\label{phistarstrong-plaw01}
\phi_{\star}=\phi_e \times \exp[-h_0\mathcal{N}].
\end{equation}

Now we can calculate the scalar spectral index and the tensor to scalar ratio as given by Eqs.~\eqref{nsasstrong} and \eqref{rstrong}, respectively. The range of the acceptable values of the free parameters of the noncomoving warm inflationary model in the  strong regime are presented in Fig.~\ref{r-nsStrongPowerlaw}. In addition, to test the warm inflationary constraint that guaranties thermal fluctuations' dominance against quantum fluctuations, we plot the $T/H$ function in Fig.~\ref{THstrongpowerlaw}.  Hence we see that the condition $T/H>1$ is satisfied for the adopted range of the model parameters. To obtain these results we have fixed the power spectrum based on its horizon exit value, i.e. $\mathcal{P}_s=2.17\times 10^{-9}$, and we have used $\zeta=3$, $a_\zeta=0.0185$, and $ b_\zeta={2.315}$ in the definition of $G(Q)$.

\subsubsection{Strong dissipative regime with generalized de Sitter scale factor}\label{SecStronga-GdSF}

  Now let us turn our attention to the generalized  de Sitter scale factors model for the strong regime.
 Following the results of the Section~\ref{Secstronga}, and the procedure of Subsection~\ref{SecStronga-powlaw}, we are going to examine the theoretical predictions for the strong dissipative regime of the noncomoving warm inflation with the generalized  de Sitter scale factors,  and compare them with the observational data sets obtained by the Planck satellite. In order to perform such a comparison,  by using Eq.~\eqref{epsilonstrong} to obtain the end of inflation values of the scalar field, we find the results presented in Table~\ref{Tab3}. Consequently, by using Eqs.~ \eqref{phi-deSitter} and  \eqref{efoldf},  the scalar field at the horizon exit for the strong regime of the warm inflationary scenario is expressed as
\begin{equation}\label{phistarstrong-INtermediate}
\phi_{\star}=\phi_e\times \exp[-h_0\mathcal{N}].
\end{equation}

By substituting Eq.\eqref{tphiweak-Interm} into Eqs.\eqref{dotphi} and \eqref{phi-deSitter},  for the strong regime with generalized de Sitter scale factors the  scalar field potential as a function of the scalar field can be expressed as
\begin{equation}\label{Potentialstrong-Interm}
U(\phi)=4 \gamma  (-\lambda+1) \left[\gamma ^2 {h_0}-(\lambda +2) \ln \left(-\frac{\lambda ^2+\lambda -2}{{\alpha}_0^2 \gamma  h_0^2 \phi ^2}\right)\right] .
\end{equation}
Also from Eqs.\eqref{Potentialstrong-Interm} and \eqref{Dissipation-Q}, for the  dissipation ratio $Q$ one obtains
\begin{equation}\label{Dissipstrong-Interm}
Q(\phi)=\frac{8 \left(\lambda ^2+\lambda -2\right)}{3 \gamma  \phi  \left(\gamma {h}_0^2 \phi ^2-8 \lambda  (\lambda +1)+16\right)}\, .
\end{equation}

To find  acceptable values of the free parameters of the model for the strong regime, one can consider the scalar spectral index and tensor to scalar ratio given by Eq.~\eqref{nsasstrong} and \eqref{rstrong}, respectively. The comparison of these quantities with the observational data is presented in Fig. \ref{r-nsStrongdeSitter}.

In addition, we test the warm inflation constraint, which guaranties the thermal fluctuations' dominance over the quantum fluctuations. To do this we plot the function $T/H$,  and we observe that the $T/H>1$ constraint is satisfied, as shown in Fig.~\ref{THstrongdeSitter}. To obtain these results we have fixed the power spectrum based on its horizon exit value, i.e., $\mathcal{P}_s=2.17\times 10^{-9}$, and have used the numerical values $\zeta=3$, $a_\zeta=0.0185,$ and $ b_\zeta={2.315}$ in the definition of $G(Q)$.

\begin{table*}[t]
  \centering
  \begin{tabular}{|c|c|c|c|c|c|c|c|c|c|c|}
\hline
\hline
$\gamma$ & $-\lambda$ & $\phi_\star$ & $\phi_{\rm e}$ & $h_0$ & ${\alpha}_0$ & $n_{\rm s}$ & $r_\star \times 10^{-17}$ & $-\mathcal{P}_{s0}\times 10^{-22}$ & $T(\phi_\star)/H(\phi_\star)$ & $\mathcal{N}$\\
\hline
$4.20$ & $18.890$ & $0.000143554$ & $0.00525699$ & $0.06001$ & $27.5$ & $0.950937$ &$6.3548$ & $9.66527$&  $1.11294$ &$60$\\
\hline
$4.25$ & $18.895$ & $0.000118045$ & $0.00432542$ & $0.06002$ & $28$ & $0.952282$ &$2.93027$ & $5.32686$&  $1.09665$ &$60$\\
   \hline
 $4.30$ & $18.897$ & $0.000119332$ & $0.00437521$ & $0.06003$ & $28.2$ & $0.953098$ &$3.06041$ & $5.592$&  $1.0807$ &$60$\\
   \hline
   $4.35$ & $18.899$ & $0.000113554$ & $0.00416586$ & $0.06004$ & $28.5$ & $0.954025$ &$2.51636$ & $4.85727$&  $1.06517$ &$60$\\
   \hline
   \hline
  \end{tabular}
\caption{Estimation of  the free parameters of the noncomoving warm inflationary model with generalized de Sitter scale factors by using the Planck 2013, 2015 and 2018 data sets, and with $C_\gamma=55$, and $n=1$, respectively.  We have restricted our analysis to  real values of the scalar field at the horizon exit. To obtain the presented results we have fixed the power spectrum to its horizon exit value, i.e. $\mathcal{P}_s=2.17\times 10^{-9}$,  and we have used  the numerical values $a_\zeta=0.0185$, and $ b_\zeta={2.315}$, respectively.}\label{Tab3}
\end{table*}

\begin{figure*}[tb]
\centering
\includegraphics[scale=.81]{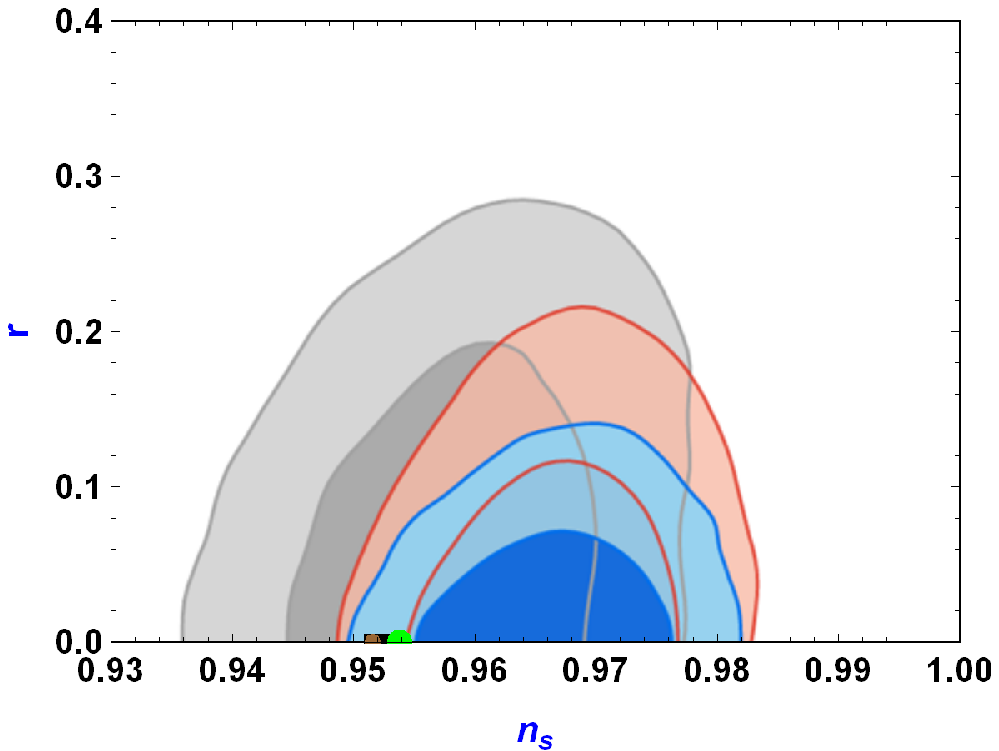}
\includegraphics[scale=.81]{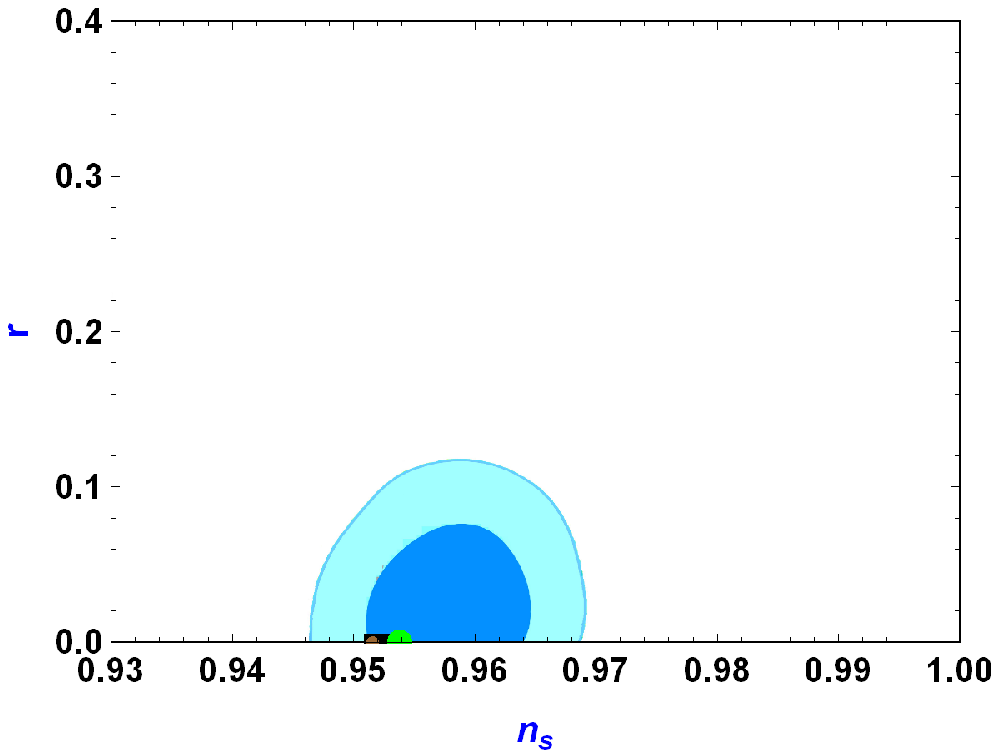}
\caption{
 The $r-n_s$ diagram comparing the predictions of the non-comoving warm inflation model in the strong dissipative regime with generalized de Sitter scale factors having the free parameters of Table~\ref{Tab3}, with the Planck $2013$, $2015$ and 2018 data sets. In the left panel, the likelihoods of  the Planck 2013 data are represented by grey contours, of the Planck TT+lowP data with red contours, while the Planck TT,TE,EE+lowP (2015) data are indicated by blue contours. In the right panel, the results of the Planck 2018 data are depicted by dark and light blue colors, respectively, corresponding to the $68\%$ and $95\%$ confidence levels, respectively. In both Figures the thick black lines indicates  the theoretical predictions of the noncomoving warm inflationary model, with the small, brown, and large, green, circles corresponding to the values of $n_{\rm s}$ at the number of e-folds $\mathcal{N}=55$ and $\mathcal{N}=65$, respectively.}
\label{r-nsStrongdeSitter}
\end{figure*}
\begin{figure}[tb]
\includegraphics[scale=.61]{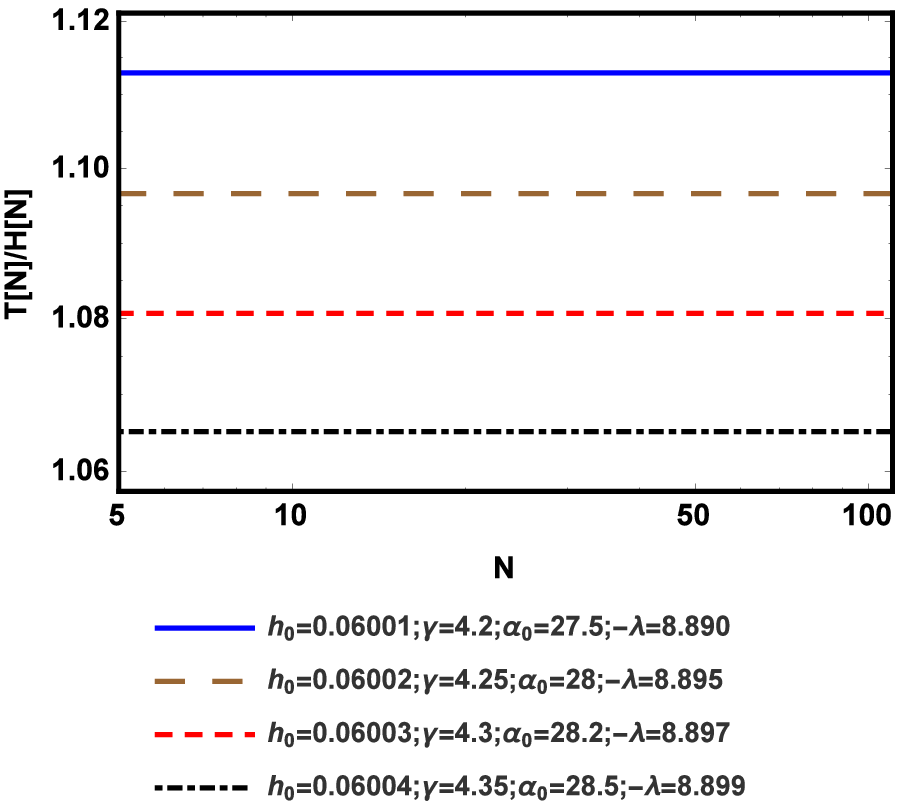}
\caption{
 The ratio of the temperature and of the Hubble parameter in the course of the inflationary evolution of the non-comoving warm inflationary model in the strong dissipative regime with generalized de Sitter scale factors, with $n=1$,  as a function of the number of e-folds, and for different numerical values of the parameters of the model.}
\label{THstrongdeSitter}
\end{figure}

\section{Discussions and final remarks}\label{sect5}

The assumption of the comoving motion of all the matter/energy components of the Universe is one of the cornerstones of present day cosmology. This mathematical choice of universal reference frame allows a simple, but powerful theoretical description of the cosmological models, and it directly leads to the Friedmann equations of standard cosmology, to the $\Lambda$CDM model, and to the inflationary paradigm, all formulated in the comoving frame. But, despite its remarkable success, still one may ask if from a physical and cosmological point of view the assumption of the existence of a universal frame is really justified, especially if one takes into account the very different nature of the major constituents of the Universe, dark energy, dark and baryonic matter? If this is indeed the case, certainly a significant amount of fine tuning would be necessary to establish the universal cosmological frame. Alternatively, one may also assume that some specific physical processes in the early Universe led to the vanishing of all forms of anisotropy, and of the possible differences in the matter and energy components.

In the present study, we have investigated the theoretical possibility that in the very early Universe, composed of a mixture
of two basic constituents, namely, scalar field and radiation,  the matter and field components have different four-velocities. More exactly, we have considered this possibility in the framework of the warm inflationary model, in which matter, originally existing in the form of radiation, is generated during the initial cosmological expansion due to the disintegration of the scalar field. The warm inflationary scenario does not require a reheating phase at the end of the inflationary era. Our basic assumption, and starting point in our analysis, is that the four velocities of the radiation and scalar field are different, and thus they are not comoving. Such an interpretation can also be supported by a simple physical argument related to the decay of particles. Let's assume that a particle of mass $M$ decays into two particles with masses $m_1$ and $m_2$. The law of conservation of energy, as applied to the system of reference in which the particle is at rest, gives $M=E_{10}+E_{20}$, where $E_{10}$ and $E_{20}$ are the energies of the emerging particles. Then by taking into account the energy conservation relation $ E_{10}^2-m_1^2=E_{20}^2-m_2^2$, one can easily obtain the energies of the decay products as $E_{10}=\left(M^2+m_1^2-m_2^2\right)/2M$, and $E_{20}=\left(M^2-m_1^2+m_2^2\right)/2M$, and obviously $E_{10}\neq E_{20}\neq M$. Consequently, the decay products will move with different four-velocities with respect to the particle generating them.

From a basic theoretical point of view the considered cosmological model with noncomoving scalar field and radiation fluid is formally equivalent to a Universe model containing a single anisotropic fluid, having different thermodynamic pressure components along the coordinate axes of the adopted coordinate system. Moreover, the equivalent thermodynamic parameters of the anisotropic system are functions of the four velocities of the two components, and of their energy densities and pressures. As a first step in our analysis we have explicitly obtained the thermodynamic parameters of the anisotropic mixture of radiation and scalar field. Since the resulting single fluid description leads to an anisotropic matter/energy distribution, the resulting geometry is also anisotropic. We have explicitly obtained the gravitational field equations describing the non-comoving scalar field  and dark radiation
mixture for the simplest homogeneous and anisotropic Universe,  described by a Bianchi type I geometry.

We have implemented the basic idea of the warm inflation by splitting the total energy conservation equation into two components, describing the decay of the scalar field, accompanied by radiation fluid creation, respectively. This splitting can be done naturally, without imposing any arbitrary functional form for the decay and creation rates, with $\Gamma _{\phi}$ and $\Gamma _{rad}$ determined by a function $F$ that depends on the differences in the four-velocities of the scalar field and radiation fluid, as well as on the energy densities and thermodynamic pressures of the constituents of the very early Universe.  In building our theoretical model we have assumed that the difference in the four-velocities
of the scalar field and radiation fluid is rather small, and therefore one can also naturally consider the alterations of the isotropic geometry as small, thus
representing just a small perturbation of the homogeneous and isotropic Friedmann-Lemaitre-Robertson-Walker type background metric. But even if there is only a slight difference between the four-velocities of the scalar field and radiation, the early Universe would achieve some anisotropic features, and its geometry will depart from the standard FLRW one. This aspect already appears in Eq.~(\ref{71}), describing the radiation generation from the scalar field, and which contains the anisotropic radiation creation term $2\alpha ^2 \dot{\phi}^2H_3$, which indicates that the distribution of the  newly produced radiation is anisotropically distributed along the coordinate axis.

One of the essential tests of any physical model is its comparison with observations. To compare the noncomoving warm inflationary model with the Planck data we have adopted the standard slow roll approximation, which is commonly used in inflationary models. In the formalism of the slow roll inflation an essential quantity is the dissipation function, which can be represented as a function of $\alpha $, $\dot{\alpha}$, and $H$, where $\alpha$ is the rotation angle in the velocity space. When $\alpha =0$, the four-velocities of the scalar field and of the radiation fluid do coincide. In the present approach we have assumed that $\alpha $ is a decreasing function of time, with $\alpha \propto t^{-h_0m}$, and $\alpha \propto e^{-h_0\gamma t}$, respectively. The parameters $h_0$ and $m$ can be determined from the observational data, and this would also fix the transition towards a universal comoving frame of the warm inflationary model. For example, in the weak dissipation regime, $m\approx 30$, and $h_0\approx 0.01$ (see Table~\ref{Tab1a}, which would give a decay law of $\alpha $ of the form $\alpha \propto t^{-0.3}$, which indicates just a slow transition towards isotropy. However, in the case of the generalized de Sitter expansion the decay of $\alpha $ is of the form $\alpha \propto e^{-3.3t}$ (see Table~\ref{Tab1b}), which would guarantee that the Universe enters in an isotropic and comoving phase at the end of the inflationary era. In the case of the strong dissipative regime $\alpha \propto t^{-0.8}$ (Table~\ref{Tab2a}) and $\alpha \propto e^{-0.24t}$ (Table~\ref{Tab3}), respectively. Hence, in the noncomoving warm inflationary model with model parameters consistent with the Planck data, the possibility that a residual anisotropy from the early Universe did survive as a consequence of the noncomoving character of the scalar field and radiation fluid evolution cannot be excluded a priori.

The present analysis, based on the standard slow roll approximation, also allows us to fix the functional form of the potential of the scalar field by essentially using the theoretical model, as well as the observations. Therefore there is no need to postulate in advance different forms of the scalar field potentials. In the case of the strong dissipative phase of noncomoving evolution with power law scale factor the potential is also of power law form, $U(\phi)\propto \phi ^{-2/h_0m}$, while in the other considered cases the potential is of a nonstandard form, involving quadratic and logarithmic terms. Finding the interpretation of these potentials from the elementary particle physics perspective is a matter of further research.

Further constraints on the noncomoving warm inflationary model can be obtained by considering the imprints of the noncomoving expansion on the Cosmic Microwave Background. The photons generated at the end of inflation at cosmological distances move in the Bianchi type I Universe, having a geometry remnant from the noncomoving motion in the early Universe, along the geodesic lines, given by
\be
\frac{du^{\mu }}{d\lambda }+\Gamma ^{\mu }_{\alpha \beta}u^{\alpha }u^{\beta }=0,
\ee
where $\lambda $ is an arbitrary affine parameter, to be determined from the solution of the geodesic equations, while the Christoffel symbols $\Gamma ^{\mu }_{\alpha \beta}$ are obtained from the metric Eq.~(\ref{7}), and they are given by $\Gamma ^{0 }_{i i}=a_i^2H_i$, and $\Gamma ^{ i}_{0 i}=H_i$, $i=1,2,3$, respectively (note that there is no summation upon $i$ in the expression of the Christoffel symbols). As for the four-velocity $u^{\mu }=dx^{\mu }/d\lambda $ of the photons, it satisfies the normalization condition $u^{\mu }u_{\mu }=0$, giving $\left(u_0\right)^2=a_i^2\left(u^i\right)^2$. The temperature of the Cosmic Microwave Background distribution  is determined by the simple relation
\be
T\left(\hat{\vec{u}}\right) =\frac{T_{*}}{1+z\left(\hat{\vec{u}}\right)},
\ee
where the redshift $z$ is defined as $1+z\left(\lambda _e\right)=\tau \left(\lambda _r\right)/\tau \left(\lambda _e\right)$, where by $\tau  \left(\lambda _r\right)$ we have denoted the time difference of the received signals. The present day values of the components of the photon four-velocity are denoted as $u^i\left(t_0\right) = \hat{u}^i$, $i=1,2,3$. $T_{*}$ denotes the last scattering temperature,  which does not depend on the direction. But due to the possible presence of anisotropies in the geometry of the Universe,  photons traveling from different directions will be redshifted by distinct  amounts. Therefore in an anisotropic Universe one must consider the  spatial average $\bar{T}$ of the temperature field, which is given by $4\pi \bar{T}=\int{T\left(\hat{\vec{u}}\right)d\Omega _{\hat{\vec{u}}}}$.  Hence  the anisotropies in the temperature field are given by $\delta T\left(\hat{\vec{u}}\right)=1-T\left(\hat{\vec{u}}\right)/\bar{T}$. Another important observational parameter that could be used to test the anisotropy survival from the early noncomoving evolution is the multipole spectrum $Q_l$,  which is obtained by considering the coefficients in the spherical expansion of the  anisotropic temperature field. The quadrupole $Q_2$ is defined as \cite{Tomi1,an1}
\bea
Q_2=\frac{2}{5\sqrt{3}}\sqrt{e_z^4+e_y^4-e_z^2e_y^2}=\frac{2}{5\sqrt{3}}e_z^2,
  \eea
where the eccentricities $e^2_y$ and $e_z^2$ are defined as $e_y^2=\left(a_1/a_2\right)^2-1$ and $ e_z^2=\left(a_1/a_3\right)^2-1$, respectively \cite{Tomi1}.

By using the values obtained from the best fit of the quadrupole measured by the Planck satellite, we can also  constrain, at least in principle,  the physical, cosmological and geometrical parameters of the noncomoving warm inflationary model. Another independent test of the model could be provided by the observations of the anisotropies in the dynamical rates describing the expansion of the Universe (the Hubble parameters), which  would lead to luminosity distance--redshift relationships that are non-invariant rotationally \cite{Tomi1}.  Such relationships obtained from the study of the Ia type supernovae may be therefore used to strongly constrain the noncomoving warm inflationary models, and to identify their imprints on the anisotropies present in the Universe.

There is some observational evidence, as shown by the recent Planck data \cite{1}-\cite{pl2018a} that suggest that even on very large cosmological scale the Universe is homogeneous and isotropic, small departures from isotropy could still be present on the cosmic scales. In the present study we have proposed a model that may account from the presence of these anisotropies as generated during the warm inflationary evolution in the very early Universe. One of the basic postulates of present day cosmology is the possibility of choosing a comoving  reference frame for all components of the Universe, no matter their nature. However, there is no general principle requiring that such a frame must have existed at all times in the Universe, and therefore in our present investigation we have considered that the inflaton scalar field generating the radiation in the very early Universe did in fact have a different four-velocity as compared to those of the photons. Hence a comoving frame could have been established dynamically, due to the time decay of the angle $\alpha$ describing the differences between the four-velocities of the two major initial constituents of the Universe, scalar field and radiation, respectively. The possibility of noncomoving cosmological motions, their influence on the evolution of the Universe, as well as the observational significance of the obtained results, will be considered in  a follow up publication.

\section*{Acknowledgments}

H. S. thanks H. Firouzjahi for constructive discussions on inflation and perturbations.

\end{document}